# Structural Symmetry, Multiplicity, and Differentiability of Eigenfrequencies

Shiyao Sun[1] and Kapil Khandelwal[2]

[1]Graduate Student, Dept. of Civil & Env. Engg. & Earth Sci., University of Notre Dame, Notre Dame, IN 46556, United States.

[2]Associate Professor, Dept. of Civil & Env. Engg. & Earth Sci., 156 Fitzpatrick Hall, University of Notre Dame, Notre Dame, IN 46556, United States, Email: kapil.khandelwal@nd.edu
ORCID: 0000-0002-5748-6019, (Corresponding Author)

## Abstract

This work investigates the multiplicity and differentiability of eigenfrequencies in structures with various symmetries. In particular, the study explores how the geometric and design variable symmetries affect the distribution of eigenvalues, distinguishing between simple and multiple eigenvalues in 3-D trusses. Moreover, this article also examines the differentiability of multiple eigenvalues under various symmetry conditions, which is crucial for gradient-based optimization. The results presented in this study show that while full symmetry ensures the differentiability of all eigenvalues, increased symmetry in optimized design, such as accidental symmetry, may lead to non-differentiable eigenvalues. Additionally, the study presents solutions using symmetric functions, demonstrating their effectiveness in ensuring differentiability in scenarios where multiple eigenvalues are non-differentiable. The study also highlights a critical insight into the differentiability criterion of symmetric functions, i.e., the completeness of eigen-clusters, which is necessary to ensure the differentiability of such functions.

# 1 Introduction

Optimizing the eigenfrequencies of structures is a highly effective approach in vibration control [1]. Vibration-related optimization applies broadly, serving both in scenarios where minimizing vibrations is required [2-5] and in contexts where maximizing oscillation amplitudes is advantageous [6-8]. One of the main challenges in eigenvalue optimization lies in the occurrence of multiple (repeated) eigenfrequencies (eigenvalues). Contrary to simple (non-repeated) eigenvalues, multiple eigenvalues may not be Fréchet differentiable [9], i.e., only Gâteaux derivatives may exist [10]. The potential lack of Fréchet differentiability of multiple eigenvalues poses additional challenges in gradient-based optimization [11]. Thus, from a practical viewpoint, it is crucial to understand the origin of such multiplicities and differentiability of multiple eigenvalues. The aim of this work is threefold: (a) to understand the source of multiple eigenvalues in structural systems, (b) to investigate the differentiability of multiple eigenvalues, and (c) to formulate *differentiable* functions of *multiple* eigenvalues that can be straightforwardly used in gradient-based optimization.

In optimization, multiple eigenvalues commonly arise due to symmetries in structures [10-12]. The significance of symmetry in structural mechanics has long been recognized. The study of symmetry in solids/structures was established within a mathematical framework in the 1980s and 1990s [13-15], and remains an active field of research. In mechanics, the understanding of symmetry is built upon the foundation of group theory and representation theory of abstract groups [16-18]. Such a group-theoretic approach provides a systematic framework to understand a structural system without any prior computational analysis [15]. For example, many studies have investigated the block diagonalization of the stiffness and mass matrices based on the group categorizations of symmetries in truss structures [15, 19] and continuum structures [20, 21]. The study of symmetry applies to vibration and stability problems, which belong to a broader category of generalized eigenvalue problems [22, 23]. In vibration [15] and stability problems concerning bifurcations [14, 19, 24], the group-theoretic approach can determine multiplicities of repeated eigenvalues without performing any eigen-analysis. Although a powerful method, it requires additional effort to identify and classify all symmetries, along with the computation of a symmetry-adapted basis that decouples the system. Nevertheless, understanding symmetry can offer valuable insights into the eigen-analysis results regarding eigenvalue multiplicities.

While extensive research is dedicated to analyzing eigenvalue problems using group theory, fewer studies have explored the impact of symmetry on eigenvalue *optimization* problems. Studying the influence of symmetry on eigenvalue optimization is essential and a logical progression, given the close relationship between structural symmetry and eigenvalue multiplicity. One goal of this work is to elucidate the effect of symmetry on the multiplicity of the eigenvalues. Rather than employing analytical group-theoretic tools, the focus is on understanding the role of the geometric and design variable symmetries on the multiplicity of the eigenvalues using illustrative examples. To this end, eigenvalues of 3-D truss systems with different symmetry configurations are numerically investigated.

As mentioned, multiple eigenvalues, particularly those not Fréchet differentiable [9], pose challenges for gradient-based eigenvalue optimization. Before computing design sensitivities, it is essential to ensure the differentiability of multiple eigenvalues. Consequently, this work also examines the differentiability of the multiple eigenvalues under different symmetry conditions. In the work by Seyranian et al. [10], the non-differentiability of multiple eigenvalues was illustrated using simple analytical examples. The authors addressed this challenge in optimization by employing the directional derivative approach when dealing with multiple eigenvalues. In contrast, in their work on the symmetry reduction method in optimization, Kosaka and Swan [25] presented a symmetry scenario where the multiple eigenvalues were differentiable. The authors recognize ambiguities in existing literature regarding the differentiability of multiple eigenvalues in design optimization and seek to identify the conditions under which multiple eigenvalues are differentiable. Thus, symmetry conditions are discussed based on the considered 3-D trusses wherein the multiple eigenvalues are differentiable. Crucially, the symmetry scenarios in which the multiple eigenvalues are not differentiable are also clarified.

To address the issue of the non-differentiability of multiple eigenvalues, existing studies in eigenvalue structural optimization generally adopt two main approaches. In the first approach, in cases where the multiple eigenvalues are only Gâteaux differentiable, directional derivatives are used to formulate the optimization problem [26, 27]. The directional derivative method is well-established and has been used in many studies [28-32]. The second approach, which has received comparatively less attention, addresses the non-differentiability of multiple eigenvalues by employing smooth symmetric functions. A symmetric function is a function that is invariant under the permutation of its arguments, and such a symmetric function of eigenvalues is differentiable [9, 33]. Symmetric functions such as the mean, $p$-norm, and Kreisselmeier-Steinhauser (KS) functions [34] are frequently used in eigenvalue topology optimization to resolve the non-differentiability of multiple eigenvalues. However, this study demonstrates that these functions require careful construction since they are not guaranteed to be differentiable when defined over an arbitrary set of eigenvalues. New findings regarding the differentiability criterion of symmetric functions and guidance on constructing such differentiable functions of multiple eigenvalues are presented, clarifying and refining the understanding of this critical topic.

The rest of the paper is organized as follows. Section 2 provides background knowledge on the eigen-analysis of truss structures in free vibration and the sensitivity analysis of simple eigenvalues. Section 3 presents a series of results that demonstrate the effect of symmetries in 3-D truss systems on the multiplicity and differentiability of eigenvalues. Section 4 concludes the results and the findings of the study.

## 2 Background

### 2.1 Generalized Eigenvalue Problem

The generalized eigenvalue problem for a 3-D truss system is given by [22]:

$$\boldsymbol{K}\boldsymbol{\phi}_q = \lambda_q \boldsymbol{M}\boldsymbol{\phi}_q, \quad q = 1, 2, \dots, n_f \tag{1}$$

where $\boldsymbol{K}$ and $\boldsymbol{M}$ are the global stiffness and mass matrices considering the free degrees of freedom, respectively; $\lambda_q \geq 0$ and $\boldsymbol{\phi}_q$ are the eigenvalues and eigenvectors of the generalized eigensystem in Eq. (1), respectively; and $n_f$ are the total number of (free) degrees of freedom in the system. The eigenvalues $\lambda_q$ are arranged in ascending order, i.e., $\lambda_1 \leq \lambda_2 \leq \cdots \leq \lambda_{n_f}$, and $\omega_q \stackrel{\text{def}}{=} \sqrt{\lambda_q}$ is the $q^{th}$ eigenfrequency of the system. Note that both $\boldsymbol{K}$ and $\boldsymbol{M}$ are symmetric matrices. Moreover, the eigenvectors are $\boldsymbol{M}$-orthonormalized, i.e., $\boldsymbol{\phi}_a^T \boldsymbol{M} \boldsymbol{\phi}_b = \delta_{ab}$ and $\boldsymbol{\phi}_a^T \boldsymbol{K} \boldsymbol{\phi}_b = \lambda_b \delta_{ab}$, where $\delta_{ab} = \begin{cases} 1 & a = b \\ 0 & a \neq b \end{cases}$ is the Kronecker delta symbol. Stiffness matrix $\boldsymbol{K}$ and mass matrix $\boldsymbol{M}$ are assembled using corresponding element stiffness ($\boldsymbol{K}^e$) and mass ($\boldsymbol{M}^e$) matrices. The element stiffness ($\boldsymbol{K}^e$) and mass ($\boldsymbol{M}^e$) matrices in the global coordinate system are transformed from the element stiffness ($\overline{\boldsymbol{K}}^e$) and mass ($\overline{\boldsymbol{M}}^e$) matrices in the local coordinate system.

Element stiffness matrix ($\overline{\boldsymbol{K}}^e$) in the local coordinate system is given by:

$$\overline{K}^e = \frac{x_e E}{L_e}\begin{bmatrix} 1 & -1 \\ -1 & 1 \end{bmatrix} \qquad (2)$$

where $x_e$ and $L_e$ are the cross-sectional area and the length of the truss element, respectively. $E$ is the Young's modulus of the material. In structural optimization, the element cross-sectional area $x_e$ is frequently considered as the element design variable. In this study, the term design variable also refers to the element cross-sectional area $x_e$.

Element mass matrix ($\overline{M}^e$) in the local coordinate system is given by:

$$\overline{M}^e = \frac{\rho_m x_e L_e}{6}\begin{bmatrix} 2 & 1 \\ 1 & 2 \end{bmatrix} \qquad (3)$$

where $\rho_m$ is the mass density of the material. To transform the element stiffness and mass matrices in the local coordinate to the global coordinate system (Figure 1), a transformation is constructed as follows:

$$T_e = \begin{bmatrix} \tau_1^e & \tau_2^e & \tau_3^e & 0 & 0 & 0 \\ 0 & 0 & 0 & \tau_1^e & \tau_2^e & \tau_3^e \end{bmatrix} \qquad (4)$$

where the scalars $\tau_1^e$, $\tau_2^e$ and $\tau_3^e$ are defined as the dot product between the unit vector of the truss element $\bar{e}$ (local axes) and the unit vectors $e_1$, $e_2$ and $e_3$ in the global coordinates, respectively, i.e.,

$$\tau_a^e = \bar{e} \cdot e_a \quad (a = 1, 2, 3) \qquad (5)$$

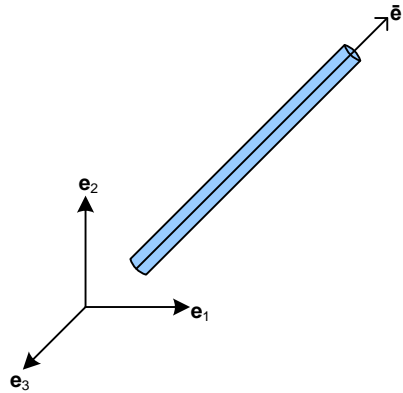

Figure 1. Truss element: local and global axes.

The element stiffness matrix $K^e$ and $M^e$ in the global system are then obtained by the following transformations:

$$K^e = T_e^T \bar{K}^e T_e$$
$$M^e = T_e^T \bar{M}^e T_e \tag{6}$$

## 2.2 Eigenvalue Sensitivity Analysis

Let $\lambda_q$ be a simple eigenvalue such that $K\phi_q = \lambda_q M\phi_q$, then the sensitivity of $\lambda_q$ w.r.t the $r^{th}$ density variable, $x_r$, is given by:

$$\frac{d\lambda_q}{dx_r} = \phi_q^T \left( \frac{dK}{dx_r} - \lambda_q \frac{dM}{dx_r} \right) \phi_q \tag{7}$$

Eq. (7) can be straightforwardly derived using direct differentiation [35]. However, Eq. (7) may not be valid when the multiplicity of the eigenvalue is greater than one because the repeated eigenvalue may not be Fréchet differentiable. The multiplicity and differentiability of eigenvalues of structures under different symmetry conditions are explored in the next section.

## 3 Results

As symmetry is the key source of multiple eigenvalues, 3-D truss systems with different geometric symmetries are investigated. To this end, 3-D dome trusses with $C_{Nv}$ symmetries and 3-D trusses with symmetries corresponding to those of platonic solids – i.e., tetrahedral ($T_d$), octahedral ($O_h$), dodecahedral ($I_h$) and icosahedral ($I_h$) symmetries – are studied. The orders of these finite groups, i.e., number of elements in these groups are: $|C_{Nv}| = 2N$, $|T_d| = 24$, $|O_h| = 48$ and $|I_h| = 120$. The symmetries of these trusses can be explained through the symmetry group of a geometric object in 3-D space or the point group [36]. The symmetry group of the set of points $P$ is the set of isometries of $\mathbb{R}^3$ that map $P$ to itself. Simply put, the isometric transformations (as elements in the symmetry group) must leave the transformed geometry indistinguishable from the original.

Such point groups define the symmetries exhibited by physical objects to a finite extent. For details, readers can refer to [17, 37] for the concepts of isometry and symmetry groups. In this study, isometries in these groups include rotations, inversions, and roto-inversions, which are characterized by a rotation about an axis followed by a reflection in a plane perpendicular to that axis. Furthermore, all reflections can be expressed in terms of roto-inversions.

A 3-D truss dome with $C_{Nv}$ geometric symmetry has $N$-fold rotation symmetries about the center vertical axis perpendicular to the plane containing all the supports. The dome truss also has $N$ reflection symmetry planes that are evenly arranged radially around the center vertical axis. The tetrahedron has 12 rotation symmetries, together with 12 roto-inversion symmetries. Thus, the tetrahedron has a total of 24 symmetries, and the symmetry group is isomorphic to the group $S_4$, i.e., $T_d \cong S_4$. The other geometries, octahedron and icosahedron/dodecahedron, have finite groups of rotation symmetries that are isomorphic to $S_4$ and $A_5$, respectively. These platonic solids also have isometries of inversion, which are isomorphic to the group $\mathbb{Z}_2$. Therefore, the full symmetry groups of octahedron and icosahedron/dodecahedron are $O_h \cong S_4 \times \mathbb{Z}_2$ and $I_h \cong A_5 \times \mathbb{Z}_2$, respectively.

To replicate the eigen-analysis results for the various trusses discussed in this paper, readers can refer to Appendix A for information on coordinate details and the assignment of design variables across different symmetry scenarios. The relevant quantities include length ($L_e$), cross-sectional area ($x_e$), mass density ($\rho_m$), and Young's modulus ($E$) and their units are m, mm$^2$, g/cm$^3$, and MPa, respectively. The mass density ($\rho_m$) and Young's modulus ($E$) remain constant throughout, with values of 0.5 g/cm$^3$ and 100 MPa, respectively. All analyses are carried out in MATLAB and the MATLAB scripts are also shared [38] that can be used by the readers to reproduce the results presented in this section.

## 3.1 Truss Domes

### 3.1.1 Truss dome without design variable symmetry

In the following example, the eigen-analysis is performed on a series of domes with $C_{Nv}$ symmetries. The domes are created by radially dividing the center into $N$ ($N = 3, 4, \ldots, 8$) numbers of subsections within the truss structures. The coordinates and dimensions details of the truss domes can be found in Appendix A.1, including both colored and grayscale figures. The domes studied are arranged in Figure 2. The red diamond shapes in the figures indicate the pin supports. The truss elements are color-coded to denote the different cross-sectional areas $x_e$'s (Figure 17(a)), which are also the design variables. The different colors of truss members in Figure 2 indicate that no design variable symmetry is enforced. That is, all design variables are different, and only geometric symmetry is present.

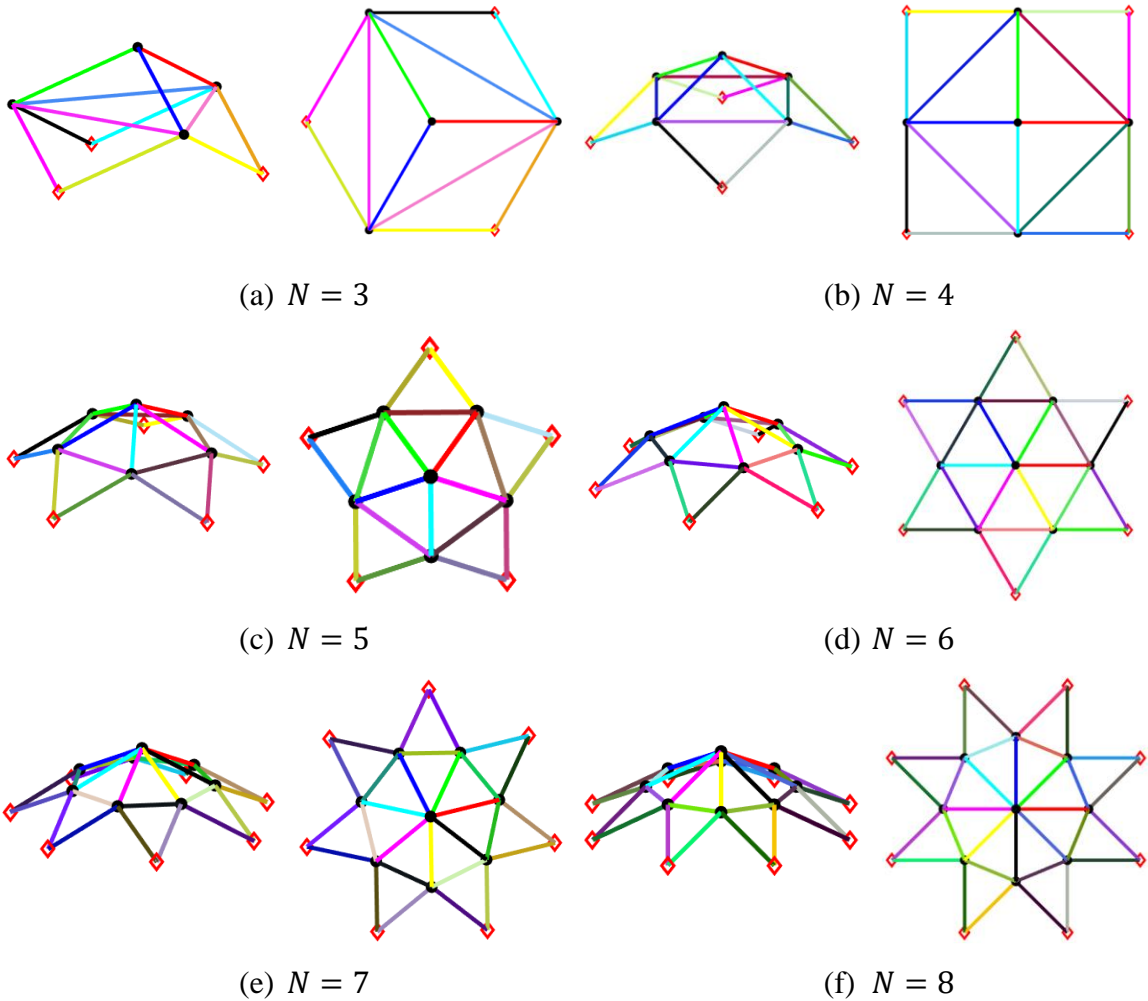

(a) $N = 3$　　　　　　　　　　　　　(b) $N = 4$

(c) $N = 5$　　　　　　　　　　　　　(d) $N = 6$

(e) $N = 7$　　　　　　　　　　　　　(f) $N = 8$

Figure 2. Truss domes with $C_{Nv}$ geometric symmetry and without design variable symmetry (left: isometric view, right: plane view).

The eigen-analysis is carried out for the domes in Figure 2, and the results are summarized in Table 1. The sensitivity check of the eigenvalues highlighted in gray in Table 1 shows that the derivatives of these eigenvalues are zero. These invariant eigenvalues arise from symmetry and are differentiable with respect to the design variables, as their derivatives equal zero. As the invariant eigenvalues are trivially differentiable, the primary focus is on the differentiability of non-invariant multiple eigenvalues.

Table 1. Eigenanalysis results of truss domes with $C_{Nv}$ geometric symmetry and without design variable symmetry.

| N | 3 | 4 | 5 | 6 | 7 | 8 |
|---|---|---|---|---|---|---|
| $\lambda_1$ | 5.287 | 14.032 | 14.928 | 10.246 | 9.450 | 7.184 |
| $\lambda_2$ | 5.948 | 14.786 | 15.787 | 16.697 | 10.117 | 10.456 |
| $\lambda_3$ | 29.758 | 16.334 | 20.404 | 17.242 | 17.801 | 10.812 |
| $\lambda_4$ | 53.231 | 37.831 | 21.288 | 23.131 | 18.209 | 18.112 |
| $\lambda_5$ | 64.715 | 46.635 | 38.637 | 23.778 | 23.541 | 18.413 |
| $\lambda_6$ | 65.697 | 59.130 | 45.396 | 34.476 | 24.010 | 22.806 |
| $\lambda_7$ | 142.781 | 60.327 | 55.426 | 48.027 | 30.736 | 23.186 |
| $\lambda_8$ | 143.258 | 120.000 | 56.872 | 53.641 | 49.885 | 27.636 |
| $\lambda_9$ | 143.545 | 173.012 | 120.000 | 55.315 | 53.196 | 51.104 |
| $\lambda_{10}$ | 279.648 | 176.405 | 120.000 | 120.000 | 55.045 | 53.464 |
| $\lambda_{11}$ | 286.808 | 177.143 | 199.160 | 120.000 | 120.000 | 55.446 |
| $\lambda_{12}$ | 354.724 | 182.938 | 204.737 | 120.000 | 120.000 | 120.000 |
| $\lambda_{13}$ | | 304.613 | 206.426 | 212.001 | 120.000 | 120.000 |
| $\lambda_{14}$ | | 311.028 | 252.162 | 217.133 | 120.000 | 120.000 |
| $\lambda_{15}$ | | 379.537 | 253.217 | 219.225 | 216.371 | 120.000 |
| $\lambda_{16}$ | | | 373.352 | 334.339 | 221.402 | 120.000 |
| $\lambda_{17}$ | | | 378.382 | 364.006 | 223.743 | 217.251 |
| $\lambda_{18}$ | | | 449.302 | 364.876 | 455.597 | 222.409 |
| $\lambda_{19}$ | | | | 497.294 | 460.058 | 225.216 |
| $\lambda_{20}$ | | | | 501.341 | 518.220 | 592.776 |
| $\lambda_{21}$ | | | | 570.280 | 520.310 | 624.747 |
| $\lambda_{22}$ | | | | | 666.027 | 626.854 |
| $\lambda_{23}$ | | | | | 669.735 | 713.473 |
| $\lambda_{24}$ | | | | | 735.523 | 716.631 |
| $\lambda_{25}$ | | | | | | 872.397 |
| $\lambda_{26}$ | | | | | | 876.232 |
| $\lambda_{27}$ | | | | | | 938.872 |

To further elucidate the effect of symmetry on the eigen spectrum, small perturbations in coordinates are added to the top node of the domes for $N = 6$ and $N = 8$ cases. As a result, the overall $C_{6v}$ and $C_{8v}$ geometric symmetries of the domes are broken (see Figure 3). The eigenvalues for these cases are presented in Table 2. The main observation is that invariant eigenvalues no longer exist, as there is no *symmetry* left in the structure. This result numerically confirms that this type of invariant eigenvalues arises from symmetry. Furthermore, all eigenvalues are now *simple,* and all the simple eigenvalues are differentiable w.r.t the design variables.

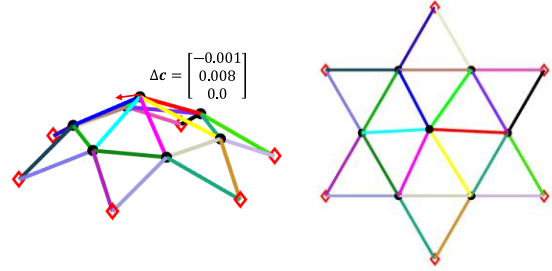

(a) $N = 6$ with top node perturbation

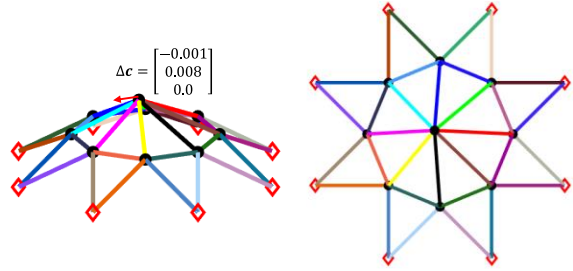

(b) $N = 8$ with top node perturbation

Figure 3. Truss domes with $C_{Nv}$ geometric symmetry and without design variable symmetry after top node perturbations (coordinate perturbations are exaggerated in figures).

Table 2. Eigenanalysis results of truss domes after top node perturbations ($C_{Nv}$ geometric symmetry and without design variable symmetry).

| N | 6 | 8 | N | 6 | 8 | N | 6 | 8 |
|---|---|---|---|---|---|---|---|---|
| $\lambda_1$ | 10.247 | 7.176 | $\lambda_{10}$ | 119.880 | 53.478 | $\lambda_{19}$ | 497.283 | 224.410 |
| $\lambda_2$ | 16.694 | 10.454 | $\lambda_{11}$ | 120.003 | 55.436 | $\lambda_{20}$ | 501.327 | 592.787 |
| $\lambda_3$ | 17.245 | 10.831 | $\lambda_{12}$ | 120.123 | 119.851 | $\lambda_{21}$ | 570.237 | 624.696 |
| $\lambda_4$ | 23.128 | 18.111 | $\lambda_{13}$ | 212.080 | 119.928 | $\lambda_{22}$ | | 626.846 |
| $\lambda_5$ | 23.779 | 18.416 | $\lambda_{14}$ | 217.222 | 120.001 | $\lambda_{23}$ | | 713.461 |
| $\lambda_6$ | 34.476 | 22.804 | $\lambda_{15}$ | 218.973 | 120.080 | $\lambda_{24}$ | | 716.580 |
| $\lambda_7$ | 48.044 | 23.183 | $\lambda_{16}$ | 334.343 | 120.138 | $\lambda_{25}$ | | 872.210 |
| $\lambda_8$ | 53.645 | 27.637 | $\lambda_{17}$ | 364.001 | 217.581 | $\lambda_{26}$ | | 876.215 |
| $\lambda_9$ | 55.305 | 51.120 | $\lambda_{18}$ | 364.868 | 222.731 | $\lambda_{27}$ | | 938.902 |

### 3.1.2 Truss dome with design variable symmetries

To study the effect of symmetries on the eigenvalue multiplicity, $C_{Nv}$ design variable symmetries are added to the truss domes (see Figure 4). Specifically, the members colored the same have the same cross-sectional area (Figure 17(b)), i.e., are assigned the same design variable. Therefore, in Figure 4, the domes exhibit both $C_{Nv}$ geometric and $C_{Nv}$ design variable symmetries. The corresponding eigen-analysis results are summarized in Table 3. It is noted that in addition to the invariant multiple eigenvalues highlighted in gray, some of the non-invariant eigenvalues cluster become repeated due to the added symmetry in the design variables.

To further investigate the differentiability of these non-invariant multiple eigenvalues w.r.t the symmetric design variables, sensitivity checks are performed on the space truss with $C_{8v}$ geometric and design variable symmetries (Figure 4(f)). The sensitivity check results comparing the sensitivities computed by the central difference method (CDM) and the analytical method are tabulated in Table 4. The perturbation in CDM used in this work is set to be 1.0E-6. The match between the sensitivities calculated by the two methods suggests that the repeated non-invariant eigenvalues are differentiable w.r.t the symmetric design variables. Note that such a match is a numerical verification of Fréchet differentiability. Recall the definition of Fréchet derivative – let $U$ be an open subset of $\mathbb{R}^n$, $\boldsymbol{x} \in U$, and $\lambda(\boldsymbol{x}): U \to \mathbb{R}$. Then $\lambda$ is Fréchet differentiable at $\boldsymbol{x}$ if there exists a ***linear*** map $L_\lambda(\boldsymbol{x}): \mathbb{R}^n \to \mathbb{R}$ such that

$$\lim_{\boldsymbol{h} \to 0} \frac{\|\lambda(\boldsymbol{x}+\boldsymbol{h}) - \lambda(\boldsymbol{x}) - L_\lambda(\boldsymbol{h})\|}{\|\boldsymbol{h}\|} = 0 \tag{8}$$

The linear map $L_\lambda(\boldsymbol{x})$ is then derivative of $\lambda(\boldsymbol{x})$ at $\boldsymbol{x}$.

To prove the differentiability of an eigenvalue, the direct approach is to write down the analytical expression of the eigenvalue $\lambda(\boldsymbol{x})$, determine the linear map $L_\lambda$ and verify that the limit defining Fréchet differentiability (Eq. (8)) holds. However, in structural eigenfrequency analysis, the direct analytical expression of eigenvalues is generally not available if the number of degrees of freedom exceeds 4, as the characteristic equation for the eigenvalue of Eq. (1) is not explicitly solvable.

Numerically, however, when the central difference approximation matches the analytical derivative (Figure 4(f)), it suggests that the linear approximation of the multiple eigenvalues provided by the analytical derivative is accurate for small perturbations around the current design point. Thus, the function behaves well enough for the linear approximation to hold, indicating Fréchet differentiability.

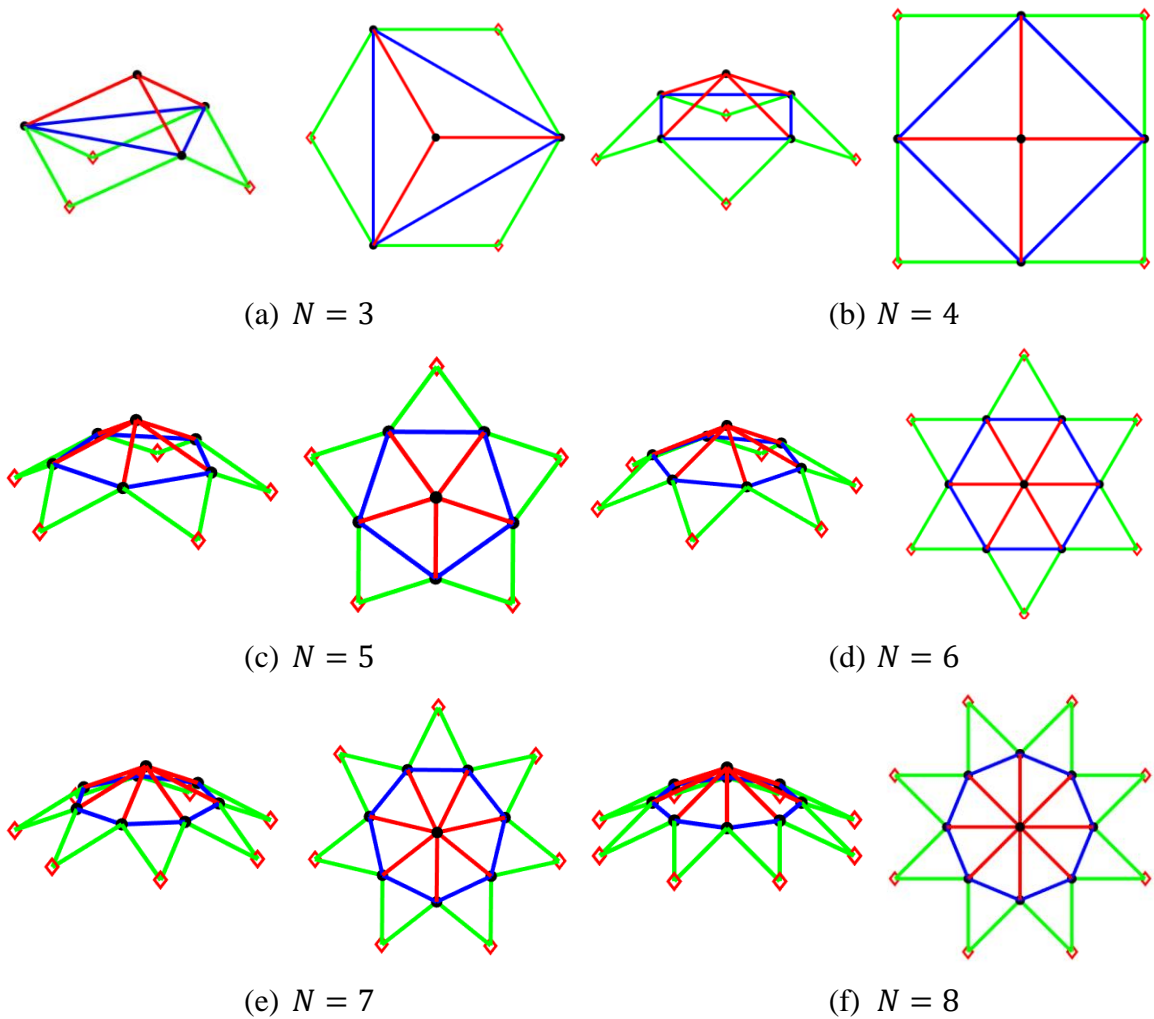

(a) $N = 3$  (b) $N = 4$

(c) $N = 5$  (d) $N = 6$

(e) $N = 7$  (f) $N = 8$

Figure 4. Truss domes with $C_{Nv}$ geometric symmetry and $C_{Nv}$ design variable symmetry (left: isometric view, right: plane view).

Table 3. Eigenanalysis results of truss domes with $C_{Nv}$ geometric symmetry and $C_{Nv}$ design variable symmetry.

| N | 3 | 4 | 5 | 6 | 7 | 8 |
|---|---|---|---|---|---|---|
| $\lambda_1$ | 2.643 | 8.533 | 15.910 | 22.932 | 28.544 | 32.639 |
| $\lambda_2$ | 2.643 | 8.533 | 15.910 | 22.932 | 28.544 | 32.639 |
| $\lambda_3$ | 11.048 | 17.327 | 24.155 | 30.034 | 34.431 | 37.508 |
| $\lambda_4$ | 96.614 | 57.498 | 56.207 | 47.294 | 45.888 | 39.886 |
| $\lambda_5$ | 97.315 | 92.559 | 56.207 | 59.063 | 45.888 | 47.550 |
| $\lambda_6$ | 97.315 | 93.319 | 88.453 | 59.063 | 61.602 | 47.550 |
| $\lambda_7$ | 125.957 | 93.319 | 89.263 | 84.551 | 61.602 | 62.762 |
| $\lambda_8$ | 125.969 | 120.000 | 89.263 | 85.402 | 80.909 | 62.762 |
| $\lambda_9$ | 125.969 | 135.904 | 120.000 | 85.402 | 81.794 | 77.530 |
| $\lambda_{10}$ | 359.419 | 136.043 | 120.000 | 120.000 | 81.794 | 78.445 |
| $\lambda_{11}$ | 359.419 | 136.043 | 151.278 | 120.000 | 120.000 | 78.445 |
| $\lambda_{12}$ | 434.150 | 136.188 | 151.930 | 120.000 | 120.000 | 120.000 |
| $\lambda_{13}$ | | 376.768 | 151.930 | 169.972 | 120.000 | 120.000 |
| $\lambda_{14}$ | | 376.768 | 156.436 | 171.616 | 120.000 | 120.000 |
| $\lambda_{15}$ | | 444.498 | 156.436 | 171.616 | 188.355 | 120.000 |
| $\lambda_{16}$ | | | 403.601 | 178.857 | 190.960 | 120.000 |
| $\lambda_{17}$ | | | 403.601 | 195.060 | 190.960 | 203.631 |
| $\lambda_{18}$ | | | 469.130 | 195.060 | 218.461 | 206.636 |
| $\lambda_{19}$ | | | | 447.106 | 218.461 | 206.636 |
| $\lambda_{20}$ | | | | 447.106 | 257.580 | 259.622 |
| $\lambda_{21}$ | | | | 513.847 | 257.580 | 280.839 |
| $\lambda_{22}$ | | | | | 515.080 | 280.839 |
| $\lambda_{23}$ | | | | | 515.080 | 346.410 |
| $\lambda_{24}$ | | | | | 584.354 | 346.410 |
| $\lambda_{25}$ | | | | | | 612.767 |
| $\lambda_{26}$ | | | | | | 612.767 |
| $\lambda_{27}$ | | | | | | 684.426 |

Table 4. Multiple eigenvalue sensitivity check results of the truss dome with $C_{8v}$ geometric symmetry and $C_{8v}$ design variable symmetry.

| CDM | Analytical | CDM | Analytical | CDM | Analytical | CDM | Analytical |
|---|---|---|---|---|---|---|---|
| $d\lambda_1/d\mathbf{x}_{sym}$ | | $d\lambda_2/d\mathbf{x}_{sym}$ | | $d\lambda_5/d\mathbf{x}_{sym}$ | | $d\lambda_6/d\mathbf{x}_{sym}$ | |
| -0.129784 | -0.129784 | -0.129784 | -0.129784 | 0.097373 | 0.097373 | 0.097373 | 0.097373 |
| 0.072012 | 0.072012 | 0.072012 | 0.072012 | 0.066802 | 0.066802 | 0.066802 | 0.066802 |
| 0.101306 | 0.101306 | 0.101306 | 0.101306 | -0.559330 | -0.559330 | -0.559329 | -0.559330 |
| $d\lambda_7/d\mathbf{x}_{sym}$ | | $d\lambda_8/d\mathbf{x}_{sym}$ | | $d\lambda_{10}/d\mathbf{x}_{sym}$ | | $d\lambda_{11}/d\mathbf{x}_{sym}$ | |
| 0.040726 | 0.040726 | 0.040726 | 0.040726 | -0.006715 | -0.006715 | -0.006715 | -0.006715 |
| 0.115301 | 0.115301 | 0.115301 | 0.115301 | 0.132824 | 0.132824 | 0.132824 | 0.132824 |
| -0.583381 | -0.583381 | -0.583381 | -0.583381 | -0.511151 | -0.511151 | -0.511151 | -0.511151 |
| $d\lambda_{18}/d\mathbf{x}_{sym}$ | | $d\lambda_{19}/d\mathbf{x}_{sym}$ | | $d\lambda_{21}/d\mathbf{x}_{sym}$ | | $d\lambda_{22}/d\mathbf{x}_{sym}$ | |
| 0.172052 | 0.172052 | 0.172052 | 0.172052 | -0.066067 | -0.066067 | -0.066067 | -0.066067 |
| -0.298264 | -0.298264 | -0.298264 | -0.298264 | -0.593324 | -0.593324 | -0.593324 | -0.593324 |
| 0.676899 | 0.676899 | 0.676899 | 0.676899 | 2.571497 | 2.571497 | 2.571497 | 2.571497 |
| $d\lambda_{23}/d\mathbf{x}_{sym}$ | | $d\lambda_{24}/d\mathbf{x}_{sym}$ | | $d\lambda_{25}/d\mathbf{x}_{sym}$ | | $d\lambda_{26}/d\mathbf{x}_{sym}$ | |
| -0.326901 | -0.326901 | -0.326901 | -0.326901 | -0.492672 | -0.492672 | -0.492673 | -0.492672 |
| -0.543397 | -0.543397 | -0.543397 | -0.543397 | -0.202622 | -0.202621 | -0.202622 | -0.202621 |
| 3.154289 | 3.154289 | 3.154289 | 3.154289 | 2.288500 | 2.288500 | 2.288500 | 2.288500 |

To further show the role of symmetry, small perturbations are again added to the top node of the domes with the $C_{6v}$ and $C_{8v}$ geometric and design variable symmetries (Figure 5) and the eigen-analysis on this perturbed system is carried out. From the eigen-analysis results summarized in Table 5, it is observed that all eigenvalues become simple. Therefore, it can be concluded that the perturbations in geometry break the overall symmetry of the structural system. As a result, neither invariant nor non-invariant multiple eigenvalues can exist anymore.

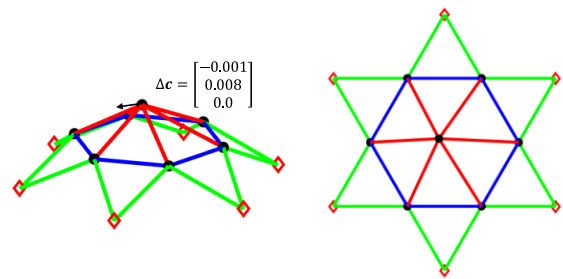

(a) $N = 6$ with top node perturbation

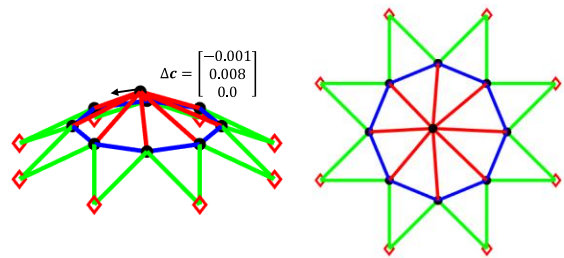

(b) $N = 8$ with top node perturbation

Figure 5. Truss domes with $C_{Nv}$ geometric symmetry and $C_{Nv}$ design variable symmetry after top node perturbations (coordinate perturbations are exaggerated in figures).

Table 5. Eigenanalysis results of truss domes with $C_{Nv}$ geometric symmetry and $C_{Nv}$ design variable symmetry after top node perturbations.

| N | 6 | 8 | N | 6 | 8 | N | 6 | 8 |
|---|---|---|---|---|---|---|---|---|
| $\lambda_1$ | 22.930 | 32.632 | $\lambda_{10}$ | 119.851 | 78.446 | $\lambda_{19}$ | 447.075 | 206.922 |
| $\lambda_2$ | 22.932 | 32.638 | $\lambda_{11}$ | 119.999 | 78.448 | $\lambda_{20}$ | 447.103 | 259.621 |
| $\lambda_3$ | 30.037 | 37.513 | $\lambda_{12}$ | 120.150 | 119.818 | $\lambda_{21}$ | 513.878 | 280.836 |
| $\lambda_4$ | 47.293 | 39.875 | $\lambda_{13}$ | 169.878 | 119.894 | $\lambda_{22}$ | | 280.838 |
| $\lambda_5$ | 59.062 | 47.549 | $\lambda_{14}$ | 171.618 | 119.999 | $\lambda_{23}$ | | 346.409 |
| $\lambda_6$ | 59.063 | 47.556 | $\lambda_{15}$ | 171.717 | 120.106 | $\lambda_{24}$ | | 346.409 |
| $\lambda_7$ | 84.549 | 62.762 | $\lambda_{16}$ | 178.856 | 120.184 | $\lambda_{25}$ | | 612.655 |
| $\lambda_8$ | 85.402 | 62.762 | $\lambda_{17}$ | 195.059 | 203.362 | $\lambda_{26}$ | | 612.768 |
| $\lambda_9$ | 85.404 | 77.528 | $\lambda_{18}$ | 195.060 | 206.639 | $\lambda_{27}$ | | 684.553 |

It is noted that when multiple eigenvalues are present, the associated eigenvectors (i.e., mode shapes) provide limited insight into the vibration behavior. This is because of the non-uniqueness of the eigenvectors corresponding to a multiple eigenvalue. While the eigenvector associated with a simple eigenvalue is unique, the eigenvectors associated with multiple eigenvalues are non-unique and span the eigenspace of those repeated eigenvalues. Because of the non-uniqueness of such eigenvectors, this study will not discuss them, focusing instead on the differentiability of multiple eigenvalues.

### *3.2 Tetrahedral Trusses*

The tetrahedral truss configuration is shown in Figure 6. At the center of the tetrahedral truss is a regular tetrahedron with four equilateral triangular faces, each vertex extending outward with an axial force element supported at its end. The coordinate and dimension details of the tetrahedral truss can be found in Appendix A.2. This tetrahedral truss structure admits both the full $T_d$ and $C_{3v}$ design variable symmetries (Figure 20). Thus, the eigen-analysis is conducted on the tetrahedral truss with the full $T_d$ (Figure 6(b)) and the reduced $C_{3v}$ (Figure 6(c)) design variable symmetries, along with the no-design variable symmetry case (Figure 6(a)).

The eigen-analysis results are summarized in Table 6. Due to the support conditions, the tetrahedral trusses have three zero eigenvalues. These rigid-body mode-related eigenvalues are omitted in the presented results. Note that the eigenvalues $\lambda_8$ and $\lambda_9$ in all the three symmetry cases are invariant multiple eigenvalues. In the no-design variable symmetry case, all non-invariant eigenvalues are simple. In the full $T_d$ symmetry case, the non-invariant multiple eigenvalues, 26.753 and 204.320, have multiplicities of 3, whereas the $C_{3v}$ design variable symmetry case has non-invariant multiple eigenvalues, 25.602 and 206.889, with multiplicities of 2. This example demonstrates that due to the symmetry reduction from $T_d$ to $C_{3v}$, the multiplicity of repeated eigenvalues is also reduced.

Next, the invariant multiple eigenvalues are always differentiable with sensitivities equal to zero. For the non-invariant multiple eigenvalues, the sensitivity check results tabulated in Table 7 ($T_d$ symmetry) and Table 8 ($C_{3v}$ symmetry) show that the eigenvalue sensitivities calculated by the analytical method consistently match those by the CDM.

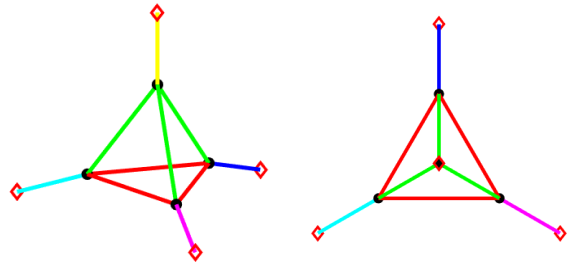

(a) No-design variable symmetry

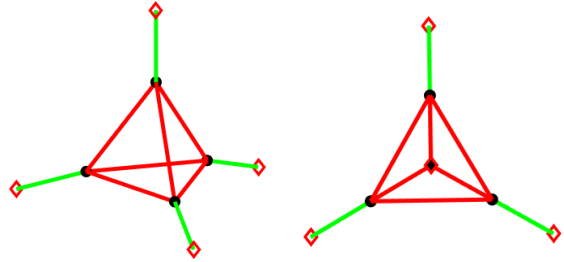

(b) Full $T_d$ symmetry

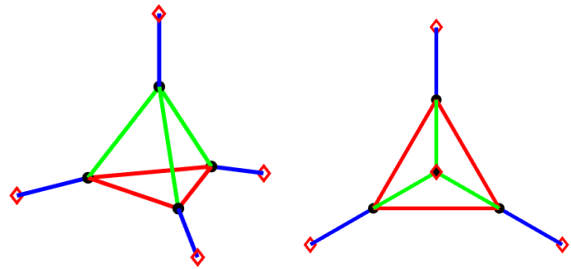

(c) $C_{3v}$ symmetry

Figure 6. Tetrahedral trusses with 3 types of symmetries (left: isometric view, right: plane view).

Table 6. Eigenanalysis results of tetrahedral trusses with 3 types of design variable symmetries.

|  | No-design var. symmetry | Full $T_d$ symmetry | $C_{3v}$ symmetry |
| --- | --- | --- | --- |
| $\lambda_1$ | 20.173 | 26.753 | 19.523 |
| $\lambda_2$ | 22.056 | 26.753 | 25.602 |
| $\lambda_3$ | 24.689 | 26.753 | 25.602 |
| $\lambda_4$ | 188.467 | 183.712 | 186.774 |
| $\lambda_5$ | 206.226 | 204.320 | 206.889 |
| $\lambda_6$ | 207.799 | 204.320 | 206.889 |
| $\lambda_7$ | 209.778 | 204.320 | 207.452 |
| $\lambda_8$ | 225.000 | 225.000 | 225.000 |
| $\lambda_9$ | 225.000 | 225.000 | 225.000 |

Table 7. Non-invariant multiple eigenvalue sensitivity check results of the tetrahedral truss with $T_d$ design variable symmetry.

| CDM | Analytical | CDM | Analytical | CDM | Analytical |
|---|---|---|---|---|---|
| $d\lambda_1/d\mathbf{x}_{sym}$ | | $d\lambda_2/d\mathbf{x}_{sym}$ | | $d\lambda_3/d\mathbf{x}_{sym}$ | |
| -0.177749 | -0.177749 | -0.177749 | -0.177749 | -0.177749 | -0.177749 |
| 0.088874 | 0.088874 | 0.088874 | 0.088874 | 0.088874 | 0.088874 |
| $d\lambda_5/d\mathbf{x}_{sym}$ | | $d\lambda_6/d\mathbf{x}_{sym}$ | | $d\lambda_7/d\mathbf{x}_{sym}$ | |
| 0.141605 | 0.141605 | 0.141605 | 0.141605 | 0.141605 | 0.141605 |
| -0.070803 | -0.070803 | -0.070803 | -0.070803 | -0.070803 | -0.070803 |

Table 8. Non-invariant multiple eigenvalue sensitivity check results of the tetrahedral truss with $C_{3v}$ design variable symmetry.

| CDM | Analytical | CDM | Analytical |
|---|---|---|---|
| $d\lambda_2/d\mathbf{x}_{sym}$ | | $d\lambda_3/d\mathbf{x}_{sym}$ | |
| -0.153134 | -0.153134 | -0.153134 | -0.153134 |
| -0.016003 | -0.016003 | -0.016003 | -0.016003 |
| 0.088569 | 0.088569 | 0.088569 | 0.088569 |
| $d\lambda_5/d\mathbf{x}_{sym}$ | | $d\lambda_6/d\mathbf{x}_{sym}$ | |
| 0.076066 | 0.076066 | 0.076066 | 0.076066 |
| 0.035969 | 0.035969 | 0.035969 | 0.035969 |
| -0.065010 | -0.065010 | -0.065010 | -0.065010 |

### 3.3 Octahedral Trusses

The octahedral truss consists of a regular octahedron in the center comprising 12 members and 6 outer elements with supports at the ends (Figure 7). The coordinates and dimensions information of the octahedral truss can be found in Appendix A.3. Aside from the full $O_h$ design variable symmetry (Figure 7(a)), the octahedral truss structure also admits $C_{4v}$ (Figure 7(b)) and $C_{2v}$ (Figure 7(c)) symmetries along with the no-design variable symmetry case (Figure 7(d)). The eigen-analysis results of the octahedral truss with the 4 types of design variable symmetries are summarized in Table 9.

Due to the support conditions, the octahedral truss has 3 zero eigenvalues which are omitted in the results. Because of the $O_h$ geometric symmetry, invariant multiple eigenvalues, i.e., $\lambda_9 = \lambda_{10} = \lambda_{11} = 1200.0$, are present in this system. Interestingly, the eigenvalues $\lambda_4 = \lambda_5 = \lambda_6 = 300.0$ are also invariant multiple eigenvalues w.r.t the symmetric design variables for $O_h$ and $C_{4v}$ symmetries. However, the invariant multiple eigenvalues ($\lambda_4 = \lambda_5 = \lambda_6 = 300.0$) split and become simple when the design variable symmetry is reduced to $C_{2v}$. Specifically, $\lambda_4$ becomes 284.330 and $\lambda_6$ becomes 317.695, while $\lambda_5$ remain invariant at 300.0. Moreover, such invariant eigenvalues no longer exist in the case with no-design variable symmetry. This finding highlights a distinction between eigenvalues invariant at 300.0 and those at 1200.0. Unlike the latter, which remain invariant with fixed geometry, the eigenvalues at 300.0 can lose their invariance when design variable symmetry is reduced, even if the geometry remains the same.

Furthermore, the multiplicity of the non-invariant multiple eigenvalues is reduced from 3 to 2 as the design variable symmetry is reduced from $O_h$ to $C_{4v}$ symmetry. Additionally, the no-design symmetry case has no non-invariant multiple eigenvalues, as expected. The sensitivity check results of the non-invariant multiple eigenvalues in the $O_h$ and $C_{4v}$ symmetry cases are summarized in Table 10 and Table 11, respectively. From Table 10 and Table 11, it can be observed that the analytical sensitivities of all multiple eigenvalues w.r.t the symmetric design variables match those approximated by CDM.

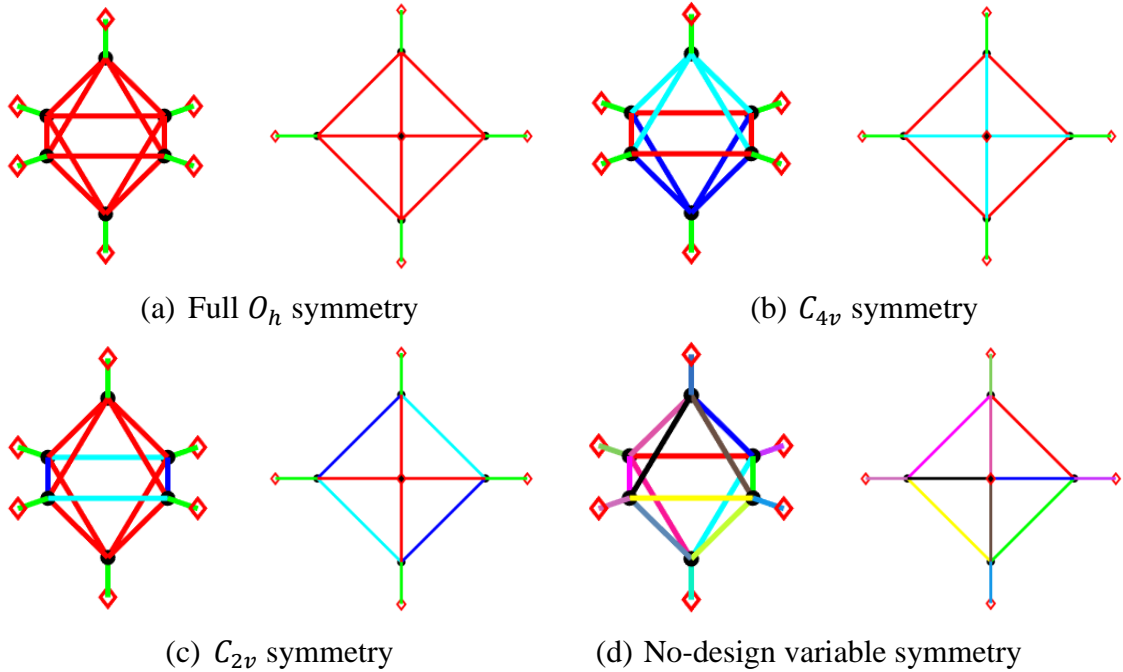

(a) Full $O_h$ symmetry    (b) $C_{4v}$ symmetry

(c) $C_{2v}$ symmetry    (d) No-design variable symmetry

Figure 7. Octahedral trusses with 3 types of symmetries (left: isometric view, right: plane view).

Table 9. Eigenanalysis results of tetrahedral trusses with 4 types of design variable symmetries.

|  | $O_h$ symmetry | $C_{4v}$ symmetry | $C_{2v}$ symmetry | No-design var. symmetry |
|---|---|---|---|---|
| $\lambda_1$ | 119.640 | 115.465 | 110.706 | 130.791 |
| $\lambda_2$ | 119.640 | 131.080 | 119.640 | 132.178 |
| $\lambda_3$ | 119.640 | 131.080 | 122.253 | 135.736 |
| $\lambda_4$ | 300.000 | 300.000 | 284.330 | 298.066 |
| $\lambda_5$ | 300.000 | 300.000 | 300.000 | 299.775 |
| $\lambda_6$ | 300.000 | 300.000 | 317.695 | 302.120 |
| $\lambda_7$ | 481.715 | 450.314 | 463.709 | 496.954 |
| $\lambda_8$ | 481.715 | 573.352 | 484.767 | 595.294 |
| $\lambda_9$ | 1200.000 | 1200.000 | 1200.000 | 1200.000 |
| $\lambda_{10}$ | 1200.000 | 1200.000 | 1200.000 | 1200.000 |
| $\lambda_{11}$ | 1200.000 | 1200.000 | 1200.000 | 1200.000 |
| $\lambda_{12}$ | 1439.042 | 1428.233 | 1424.405 | 1469.878 |
| $\lambda_{13}$ | 1439.042 | 1470.485 | 1438.297 | 1473.675 |
| $\lambda_{14}$ | 1439.042 | 1470.485 | 1439.042 | 1484.799 |
| $\lambda_{15}$ | 1584.453 | 1606.134 | 1577.877 | 1637.621 |

Table 10. Non-invariant multiple eigenvalue sensitivity check results of the octahedral truss with full $O_h$ design variable symmetry.

| CDM | Analytical | CDM | Analytical | CDM | Analytical | CDM | Analytical |
|---|---|---|---|---|---|---|---|
| $d\lambda_1/d\mathbf{x}_{sym}$ | | $d\lambda_2/d\mathbf{x}_{sym}$ | | $d\lambda_3/d\mathbf{x}_{sym}$ | | $d\lambda_7/d\mathbf{x}_{sym}$ | |
| -0.496927 | -0.496927 | -0.496927 | -0.496927 | -0.496927 | -0.496927 | -2.028868 | -2.028868 |
| 0.372696 | 0.372696 | 0.372695 | 0.372696 | 0.372695 | 0.372696 | 1.521651 | 1.521651 |
| $d\lambda_8/d\mathbf{x}_{sym}$ | | $d\lambda_{12}/d\mathbf{x}_{sym}$ | | $d\lambda_{13}/d\mathbf{x}_{sym}$ | | $d\lambda_{14}/d\mathbf{x}_{sym}$ | |
| -2.028868 | -2.028868 | -1.322495 | -1.322495 | -1.322494 | -1.322495 | -1.322495 | -1.322495 |
| 1.521651 | 1.521651 | 0.991871 | 0.991871 | 0.991871 | 0.991871 | 0.991871 | 0.991871 |

Table 11. Non-invariant multiple eigenvalue sensitivity check results of the octahedral truss with $C_{4v}$ design variable symmetry.

| CDM | Analytical | CDM | Analytical | CDM | Analytical | CDM | Analytical |
|---|---|---|---|---|---|---|---|
| $d\lambda_2/d\mathbf{x}_{sym}$ | | $d\lambda_3/d\mathbf{x}_{sym}$ | | $d\lambda_{13}/d\mathbf{x}_{sym}$ | | $d\lambda_{14}/d\mathbf{x}_{sym}$ | |
| -0.198025 | -0.198025 | -0.198024 | -0.198025 | -0.562135 | -0.562136 | -0.562135 | -0.562136 |
| 0.255348 | 0.255347 | 0.255348 | 0.255347 | 0.724859 | 0.724859 | 0.724859 | 0.724859 |
| -0.109709 | -0.109709 | -0.109709 | -0.109709 | -0.311432 | -0.311432 | -0.311433 | -0.311432 |
| -0.088864 | -0.088864 | -0.088864 | -0.088864 | -0.252260 | -0.252260 | -0.252261 | -0.252260 |

## *3.4 Dodecahedral Trusses*

Numerical examples have demonstrated that multiple eigenvalues, invariant or not, have consistent results (CDM and analytical) regarding sensitivities w.r.t to the symmetric design variables of the corresponding enforced symmetry. In this example, a truss exhibiting the geometric symmetry of a regular dodecahedron is presented to further illustrate the effect of symmetry on eigenvalues. The details about the coordinates and dimensions of the dodecahedral truss can be found in Appendix A.4. Figure 8 illustrates 2 types of admissible symmetries of the dodecahedral truss: full $I_h$ and the reduced $C_{5v}$ design variable symmetries. Table 12 records the eigenvalues of dodecahedral trusses with the 2 types of design variable symmetries considered. The first 11 zero eigenvalues corresponding to rigid body modes are omitted. In this symmetry case, invariant multiple eigenvalues of 1570.820 with a multiplicity of 10 are observed. Again, the multiplicities of the non-invariant multiple eigenvalues are reduced with the decrease in design variable symmetry. The sensitivity check results presented in Table 13 ($I_h$ symmetry) and Table 14 ($C_{5v}$ symmetry) show that these non-invariant multiple eigenvalues are differentiable w.r.t the symmetric design variables. For illustration, only the sensitivities of the first eigenvalue in each cluster of repeated eigenvalues are shown. Thus far, the eigen-analysis and sensitivity results agree with the findings in previous examples.

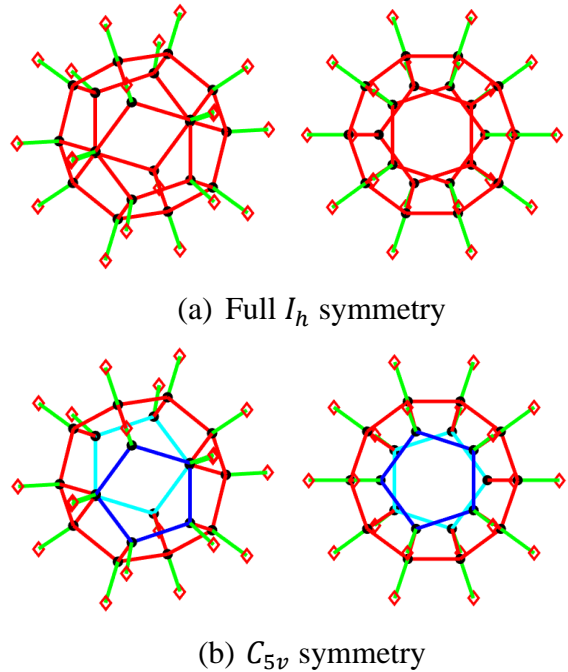

(a) Full $I_h$ symmetry

(b) $C_{5v}$ symmetry

Figure 8. Dodecahedral trusses with 2 types of symmetries (left: isometric view, right: plane view).

Table 12. Eigenanalysis results of dodecahedral trusses with 2 types of design variable symmetries.

|  | $I_h$ symmetry | $C_{5v}$ symmetry |  | $I_h$ symmetry | $C_{5v}$ symmetry |  | $I_h$ symmetry | $C_{5v}$ symmetry |
|---|---|---|---|---|---|---|---|---|
| $\lambda_1$ | 55.900 | 39.458 | $\lambda_{18}$ | 503.063 | 487.899 | $\lambda_{35}$ | 1252.258 | 1321.505 |
| $\lambda_2$ | 55.900 | 39.458 | $\lambda_{19}$ | 503.063 | 512.523 | $\lambda_{36}$ | 1252.258 | 1321.505 |
| $\lambda_3$ | 55.900 | 79.941 | $\lambda_{20}$ | 892.457 | 903.067 | $\lambda_{37}$ | 1310.443 | 1327.862 |
| $\lambda_4$ | 161.803 | 129.707 | $\lambda_{21}$ | 907.427 | 912.444 | $\lambda_{38}$ | 1310.443 | 1327.862 |
| $\lambda_5$ | 161.803 | 129.707 | $\lambda_{22}$ | 907.427 | 912.444 | $\lambda_{39}$ | 1310.443 | 1370.227 |
| $\lambda_6$ | 161.803 | 139.313 | $\lambda_{23}$ | 907.427 | 943.040 | $\lambda_{40}$ | 1570.820 | 1570.820 |
| $\lambda_7$ | 161.803 | 139.313 | $\lambda_{24}$ | 944.074 | 943.040 | $\lambda_{41}$ | 1570.820 | 1570.820 |
| $\lambda_8$ | 161.803 | 201.260 | $\lambda_{25}$ | 944.074 | 966.368 | $\lambda_{42}$ | 1570.820 | 1570.820 |
| $\lambda_9$ | 262.243 | 201.260 | $\lambda_{26}$ | 944.074 | 989.799 | $\lambda_{43}$ | 1570.820 | 1570.820 |
| $\lambda_{10}$ | 262.243 | 202.058 | $\lambda_{27}$ | 944.074 | 989.799 | $\lambda_{44}$ | 1570.820 | 1570.820 |

| | | | | | | | |
|---|---|---|---|---|---|---|---|
| $\lambda_{11}$ | 262.243 | 268.941 | $\lambda_{28}$ | 944.074 | 991.714 | $\lambda_{45}$ | 1570.820 | 1570.820 |
| $\lambda_{12}$ | 262.243 | 268.941 | $\lambda_{29}$ | 994.643 | 991.714 | $\lambda_{46}$ | 1570.820 | 1570.820 |
| $\lambda_{13}$ | 479.998 | 398.301 | $\lambda_{30}$ | 994.643 | 1022.921 | $\lambda_{47}$ | 1570.820 | 1570.820 |
| $\lambda_{14}$ | 479.998 | 398.301 | $\lambda_{31}$ | 994.643 | 1081.442 | $\lambda_{48}$ | 1570.820 | 1570.820 |
| $\lambda_{15}$ | 479.998 | 449.992 | $\lambda_{32}$ | 994.643 | 1081.442 | $\lambda_{49}$ | 1570.820 | 1570.820 |
| $\lambda_{16}$ | 479.998 | 449.992 | $\lambda_{33}$ | 1252.258 | 1261.584 | | | |
| $\lambda_{17}$ | 503.063 | 487.899 | $\lambda_{34}$ | 1252.258 | 1261.584 | | | |

Table 13. Non-invariant multiple eigenvalue sensitivity check results of the dodecahedral truss with full $I_h$ design variable symmetry.

| CDM | Analytical | CDM | Analytical | CDM | Analytical |
|---|---|---|---|---|---|
| $d\lambda_1/d\mathbf{x}_{sym}$ | | $d\lambda_4/d\mathbf{x}_{sym}$ | | $d\lambda_9/d\mathbf{x}_{sym}$ | |
| -0.128916 | -0.128916 | -0.434554 | -0.434554 | -0.795311 | -0.795311 |
| 0.064458 | 0.064458 | 0.217277 | 0.217277 | 0.397656 | 0.397655 |
| $d\lambda_{13}/d\mathbf{x}_{sym}$ | | $d\lambda_{17}/d\mathbf{x}_{sym}$ | | $d\lambda_{21}/d\mathbf{x}_{sym}$ | |
| -1.591204 | -1.591204 | -1.635255 | -1.635254 | 0.916405 | 0.916405 |
| 0.795602 | 0.795602 | 0.817628 | 0.817627 | -0.458202 | -0.458203 |
| $d\lambda_{24}/d\mathbf{x}_{sym}$ | | $d\lambda_{29}/d\mathbf{x}_{sym}$ | | $d\lambda_{33}/d\mathbf{x}_{sym}$ | |
| 1.127813 | 1.127813 | 1.328181 | 1.328180 | 1.212328 | 1.212328 |
| -0.563907 | -0.563907 | -0.664090 | -0.664090 | -0.606164 | -0.606164 |
| $d\lambda_{37}/d\mathbf{x}_{sym}$ | | | | | |
| 1.038750 | 1.038750 | | | | |
| -0.519375 | -0.519375 | | | | |

Table 14. Non-invariant multiple eigenvalue sensitivity check results of the dodecahedral truss with $C_{5v}$ design variable symmetry.

| CDM | Analytical | CDM | Analytical | CDM | Analytical |
|---|---|---|---|---|---|
| $d\lambda_1/d\mathbf{x}_{sym}$ | | $d\lambda_4/d\mathbf{x}_{sym}$ | | $d\lambda_6/d\mathbf{x}_{sym}$ | |
| 0.027138 | 0.027137 | -0.073142 | -0.073142 | -0.159129 | -0.159129 |
| 0.067491 | 0.067491 | 0.228922 | 0.228922 | 0.258677 | 0.258677 |
| -0.038531 | -0.038531 | -0.069765 | -0.069765 | -0.064336 | -0.064336 |
| -0.035597 | -0.035597 | -0.076464 | -0.076464 | -0.053561 | -0.053561 |
| $d\lambda_8/d\mathbf{x}_{sym}$ | | $d\lambda_{11}/d\mathbf{x}_{sym}$ | | $d\lambda_{13}/d\mathbf{x}_{sym}$ | |
| 0.032367 | 0.032367 | -0.638969 | -0.638969 | -0.511747 | -0.511747 |
| 0.340533 | 0.340533 | 0.487134 | 0.487134 | 0.834194 | 0.834194 |
| -0.166454 | -0.166454 | -0.003706 | -0.003706 | -0.178290 | -0.178290 |
| -0.135565 | -0.135565 | -0.002990 | -0.002990 | -0.199846 | -0.199847 |
| $d\lambda_{15}/d\mathbf{x}_{sym}$ | | $d\lambda_{17}/d\mathbf{x}_{sym}$ | | $d\lambda_{21}/d\mathbf{x}_{sym}$ | |
| -0.773699 | -0.773699 | -1.194711 | -1.194711 | 0.772652 | 0.772652 |
| 0.881391 | 0.881391 | 0.866530 | 0.866530 | -0.612132 | -0.612132 |
| -0.174234 | -0.174234 | 0.013516 | 0.013517 | 0.016240 | 0.016240 |
| -0.084083 | -0.084083 | 0.011438 | 0.011438 | 0.011498 | 0.011498 |
| $d\lambda_{23}/d\mathbf{x}_{sym}$ | | $d\lambda_{26}/d\mathbf{x}_{sym}$ | | $d\lambda_{28}d\mathbf{x}_{sym}$ | |
| 1.136256 | 1.136256 | 0.600748 | 0.600748 | 1.364453 | 1.364453 |
| -0.564129 | -0.564129 | -0.768652 | -0.768652 | -0.670935 | -0.670935 |
| -0.001856 | -0.001856 | 0.163163 | 0.163163 | -0.005264 | -0.005264 |
| -0.001529 | -0.001529 | 0.107625 | 0.107625 | -0.004295 | -0.004296 |
| $d\lambda_{31}/d\mathbf{x}_{sym}$ | | $d\lambda_{33}/d\mathbf{x}_{sym}$ | | $d\lambda_{35}/d\mathbf{x}_{sym}$ | |
| 0.359884 | 0.359884 | 0.757475 | 0.757475 | 0.272767 | 0.272767 |
| -0.832266 | -0.832266 | -0.650972 | -0.650973 | -0.574797 | -0.574798 |
| 0.238616 | 0.238616 | 0.039773 | 0.039773 | 0.157027 | 0.157027 |

| 0.235127 | 0.235128 | 0.757475 | 0.757475 | 0.154854 | 0.154854 |
|---|---|---|---|---|---|
| $d\lambda_{37}/d\mathbf{x}_{sym}$ | | | | | |
| 0.567561 | 0.567561 | | | | |
| -0.537499 | -0.537499 | | | | |
| 0.049845 | 0.049844 | | | | |
| 0.044602 | 0.044602 | | | | |

***Remarks:*** In eigenvalue optimization problems where full symmetry is enforced, the space of the design variables is reduced to the space of the symmetric design variables. These symmetric design variables are then the free variables that are optimized. The examples shown so far demonstrate a consistent match of multiple eigenvalue sensitivities computed by CDM approximation and analytical method, indicating differentiability. The following example illustrates a symmetry scenario in which multiple eigenvalues are not differentiable.

### 3.4.1 Accidental Symmetry and Non-differentiability of Multiple Eigenvalues

So far, the presented examples have explored the multiplicity and differentiability of eigenvalues when the resulting design variable symmetry is the same as the enforced symmetry. In structural optimization, the increase in design variable symmetry can also occur. The increase in symmetry occurs when the optimization process achieves a design with a higher symmetry than the initially enforced symmetry. This type of increase in symmetry is termed *accidental* symmetry in this study. The following example demonstrates accidental symmetry and highlights the challenges it brings in terms of the differentiability of multiple eigenvalues.

Table 15 tabulates the values of symmetric design variables in the full $I_h$, $C_{5v}$, and $C_{5v}$ to accidental $I_h$ symmetry conditions. The colors labeled in the table correspond to the colors of the truss elements in Figure 8. Note that in the $I_h$ and the $C_{5v}$ symmetry cases, each symmetric design variable is assigned a different value. However, in the $C_{5v}$ to accidental $I_h$ symmetry case, the symmetric design variables $x_{1_{sym}}$, $x_{3_{sym}}$ and $x_{4_{sym}}$ are assigned the same value of 100.0. With this assignment, the truss now has $I_h$ design variable symmetry even though the enforced symmetry is $C_{5v}$. Consequently, the dodecahedral truss with the $C_{5v}$ to accidental $I_h$ symmetry has the same eigenvalues as the full $I_h$ symmetry case tabulated in Table 12. This accidental symmetry corresponds to a scenario in optimization where the optimized structure exhibits higher design variable symmetry than what is explicitly enforced.

The sensitivity check results for this accidental symmetry case are summarized in Table 16. In this case, the analytical sensitivities no longer match those computed by CDM. Indeed, both the analytical and CDM sensitivities are *incorrect*, as eigenvalues are no longer differentiable. This result shows that with the accidental $I_h$ symmetry, the multiple eigenvalues are not differentiable w.r.t the symmetric design variables corresponding to the enforced $C_{5v}$ symmetry. Interestingly, in the $C_{5v}$ to accidental $I_h$ symmetry case, the union of $x_{1_{sym}}$, $x_{3_{sym}}$ and $x_{4_{sym}}$ together forms $x_{1_{sym}}$ in the $I_h$ symmetry and the multiple eigenvalues are not differentiable w.r.t $x_{1_{sym}}$, $x_{3_{sym}}$ and $x_{4_{sym}}$. On the other hand, $x_{2_{sym}}$ in the $C_{5v}$ to accidental $I_h$ symmetry case represents the same members as the $x_{2_{sym}}$ in the full $I_h$ symmetry and the multiple eigenvalues are differentiable w.r.t $x_{2_{sym}}$.

Table 15. Values of symmetric design variables in 3 symmetry conditions.

| $I_h$ symmetry | | $C_{5v}$ symmetry | | $C_{5v}$ to acc. $I_h$ symmetry | |
|---|---|---|---|---|---|
| $x_{1_{sym}}$ (red) | 100 | $x_{1_{sym}}$ (red) | 100 | $x_{1_{sym}}$ (red) | 100 |
| $x_{2_{sym}}$ (green) | 200 | $x_{2_{sym}}$ (green) | 200 | $x_{2_{sym}}$ (green) | 200 |
| | | $x_{3_{sym}}$ (blue) | 225 | $x_{3_{sym}}$ (blue) | 100 |
| | | $x_{4_{sym}}$ (cyan) | 250 | $x_{4_{sym}}$ (cyan) | 100 |

Table 16. Non-invariant multiple eigenvalue sensitivity check results of the dodecahedral truss with $C_{5v}$ to accidental $I_h$ design variable symmetry ($\boldsymbol{x}_{sym} = [x_{1_{sym}}, x_{2_{sym}}, x_{3_{sym}}, x_{4_{sym}}]$).

| CDM | Analytical | CDM | Analytical | CDM | Analytical |
|---|---|---|---|---|---|
| $d\lambda_1/d\mathbf{x}_{sym}$ | | $d\lambda_4/d\mathbf{x}_{sym}$ | | $d\lambda_9/d\mathbf{x}_{sym}$ | |
| -0.170032 | 0.079902 | -0.520175 | -0.518937 | -0.530207 | -0.287224 |
| 0.064458 | 0.064458 | 0.217277 | 0.217277 | 0.397656 | 0.397655 |
| 0.020558 | -0.104409 | 0.042811 | 0.042191 | -0.132552 | -0.254043 |
| 0.020558 | -0.104409 | 0.042811 | 0.042191 | -0.132552 | -0.254043 |
| $d\lambda_{13}/d\mathbf{x}_{sym}$ | | $d\lambda_{17}/d\mathbf{x}_{sym}$ | | $d\lambda_{21}/d\mathbf{x}_{sym}$ | |
| -1.060802 | -1.576603 | -1.255795 | -0.822979 | 0.515414 | 0.272179 |
| 0.795602 | 0.795602 | 0.817628 | 0.817627 | -0.458202 | -0.458203 |
| -0.265201 | -0.007300 | -0.189730 | -0.406138 | 0.200496 | 0.322113 |
| -0.265200 | -0.007300 | -0.189729 | -0.406138 | 0.200496 | 0.322113 |
| $d\lambda_{24}/d\mathbf{x}_{sym}$ | | $d\lambda_{29}/d\mathbf{x}_{sym}$ | | $d\lambda_{33}/d\mathbf{x}_{sym}$ | |
| 0.816407 | 0.749721 | 0.885454 | 1.044000 | 0.808218 | 0.771361 |
| -0.563907 | -0.563907 | -0.664090 | -0.664090 | -0.606164 | -0.606164 |
| 0.155703 | 0.189046 | 0.221364 | 0.142090 | 0.202055 | 0.220483 |
| 0.155703 | 0.189046 | 0.221364 | 0.142090 | 0.202055 | 0.220483 |
| $d\lambda_{37}/d\mathbf{x}_{sym}$ | | | | | |
| 0.647616 | 0.655684 | | | | |
| -0.519375 | -0.519375 | | | | |
| 0.195566 | 0.191533 | | | | |
| 0.195567 | 0.191533 | | | | |

### *3.4.2 Solution: Symmetric Function (Mean Function)*

When such an accidental symmetry occurs, the multiple eigenvalues are no longer differentiable. In this case, sensitivities w.r.t the design variables are unavailable, and alternative optimization approaches must be employed. For instance, methods based on non-smooth optimization techniques [39, 40] or on directional derivatives [26, 27] can be used. However, with these methods, the optimization process becomes more involved. A more straightforward approach to address the non-differentiability of multiple eigenvalues, which applies to many problems, is using *symmetric* functions. A symmetric smooth function of the eigenvalues in the spectrum – regardless of their multiplicities – is differentiable [9, 33]. Thus, the non-differentiability issue in the case of accidental symmetry can be resolved by employing such symmetric functions of multiple eigenvalues in the formulation of eigenvalue optimization problems.

For instance, the mean function is a symmetric (smooth) function of eigenvalues. The utilization of the cluster mean to address the non-differentiability of multiple eigenvalues was first introduced by Zhang et al. [41] in buckling-constrained topology optimization. To construct the mean function, the eigenvalues in Eq. (1) are first clustered according to their multiplicities, i.e., $\underbrace{\lambda_1 = \cdots = \lambda_{N_1}}_{\Lambda_1} < \underbrace{\lambda_{N_1+1} = \cdots = \lambda_{N_1+N_2}}_{\Lambda_2} < \cdots < \underbrace{\lambda_{N_1+\cdots+N_{n_c-1}+1} = \cdots = \lambda_{N_1+\cdots+N_{n_c}}}_{\Lambda_{n_c}}$. The mean function $\overline{\Lambda}_q$ of the $q^{th}$ cluster with $N_q$ repeated eigenvalues, which is a symmetric function of the eigenvalues, is given as

$$\overline{\Lambda}_q = \frac{1}{N_q} \sum_{\lambda_k \in \Lambda_q} \lambda_k \tag{9}$$

Even when the individual eigenvalue, $\lambda_k$, in the cluster is not differentiable, the cluster mean function $\overline{\Lambda}_q$ in Eq. (9) is a differentiable function [9, 33], i.e.,

$$\frac{d\overline{\Lambda}_q}{dx_r} = \frac{1}{N_q} \left( \sum_{\lambda_k \in \Lambda_q} \left[ \boldsymbol{\phi}_k^T \left( \frac{d\boldsymbol{K}}{dx_r} - \lambda_k \frac{d\boldsymbol{M}}{dx_r} \right) \boldsymbol{\phi}_k \right] \right) \tag{10}$$

is well-defined.

In the case of the dodecahedral truss with $C_{5v}$ to accidental $I_h$ symmetry, the eigen-clusters are constructed according to the eigen-analysis results recorded in Table 12 ($I_h$ symmetry): $\underbrace{\lambda_1 = \lambda_2 = \lambda_3}_{\Lambda_1} < \underbrace{\lambda_4 = \lambda_5 = \lambda_6 = \lambda_7 = \lambda_8}_{\Lambda_2} < \underbrace{\lambda_9 = \lambda_{10} = \lambda_{11} = \lambda_{12}}_{\Lambda_3} < \underbrace{\lambda_{13} = \lambda_{14} = \lambda_{15} = \lambda_{16}}_{\Lambda_4} < \underbrace{\lambda_{17} = \lambda_{18} = \lambda_{19}}_{\Lambda_5} < \underbrace{\lambda_{20}}_{\Lambda_6} < \underbrace{\lambda_{21} = \lambda_{22} = \lambda_{23}}_{\Lambda_7} < \underbrace{\lambda_{24} = \lambda_{25} = \lambda_{26} = \lambda_{27} = \lambda_{28}}_{\Lambda_8} < \underbrace{\lambda_{29} = \lambda_{30} = \lambda_{31} = \lambda_{32}}_{\Lambda_9} < \underbrace{\lambda_{33} = \lambda_{34} = \lambda_{35} = \lambda_{36}}_{\Lambda_{10}} < \underbrace{\lambda_{37} = \lambda_{38} = \lambda_{39}}_{\Lambda_{11}} < \underbrace{\lambda_{40} = \cdots = \lambda_{49}}_{\Lambda_{12}}$. Notice that except for $\Lambda_6$ which contains simple eigenvalue, all other eigen-clusters have multiple eigenvalues with $\Lambda_{12}$ being the invariant eigen-cluster. The sensitivity check results of the mean of all eigen-clusters in the case of accidental symmetry are summarized in Table 17. The results demonstrate that even if individual multiple eigenvalues are not differentiable (Table 16), the mean function comprising these eigenvalues remains differentiable.

Table 17. Cluster mean sensitivity check results of the dodecahedral truss with $C_{5v}$ to accidental $I_h$ design variable symmetry.

| CDM | Analytical | CDM | Analytical | CDM | Analytical |
|---|---|---|---|---|---|
| $d\bar{\Lambda}_1/d\mathbf{x}_{sym}$ | | $d\bar{\Lambda}_2/d\mathbf{x}_{sym}$ | | $d\bar{\Lambda}_3/d\mathbf{x}_{sym}$ | |
| -0.085944 | -0.085944 | -0.289702 | -0.289702 | -0.530207 | -0.530207 |
| 0.064458 | 0.064458 | 0.217277 | 0.217277 | 0.397656 | 0.397655 |
| -0.021486 | -0.021486 | -0.072426 | -0.072426 | -0.132552 | -0.132552 |
| -0.021486 | -0.021486 | -0.072425 | -0.072426 | -0.132552 | -0.132552 |
| $d\bar{\Lambda}_4/d\mathbf{x}_{sym}$ | | $d\bar{\Lambda}_5/d\mathbf{x}_{sym}$ | | $d\bar{\Lambda}_6/d\mathbf{x}_{sym}$ | |
| -1.060803 | -1.060803 | -1.090170 | -1.090170 | 0.542447 | 0.542447 |
| 0.795602 | 0.795602 | 0.817627 | 0.817627 | -0.406834 | -0.406835 |
| -0.265201 | -0.265201 | -0.272542 | -0.272542 | 0.135612 | 0.135612 |
| -0.265200 | -0.265201 | -0.272542 | -0.272542 | 0.135612 | 0.135612 |
| $d\bar{\Lambda}_7/d\mathbf{x}_{sym}$ | | $d\bar{\Lambda}_8/d\mathbf{x}_{sym}$ | | $d\bar{\Lambda}_9/d\mathbf{x}_{sym}$ | |
| 0.610937 | 0.610937 | 0.751876 | 0.751876 | 0.885454 | 0.885454 |
| -0.458202 | -0.458203 | -0.563907 | -0.563907 | -0.664090 | -0.664090 |
| 0.152735 | 0.152734 | 0.187969 | 0.187969 | 0.221363 | 0.221363 |
| 0.152734 | 0.152734 | 0.187969 | 0.187969 | 0.221364 | 0.221363 |
| $d\bar{\Lambda}_{10}/d\mathbf{x}_{sym}$ | | $d\bar{\Lambda}_{11}/d\mathbf{x}_{sym}$ | | $d\bar{\Lambda}_{12}/d\mathbf{x}_{sym}$ | |
| 0.808218 | 0.808219 | 0.692500 | 0.692500 | -3.41E-07 | 9.39E-16 |
| -0.606164 | -0.606164 | -0.519375 | -0.519375 | 6.82E-07 | -1.11E-17 |
| 0.202055 | 0.202055 | 0.173124 | 0.173125 | 1.14E-07 | -2.53E-16 |
| 0.202055 | 0.202055 | 0.173125 | 0.173125 | -1.14E-07 | -2.11E-16 |

*3.4.3 Example – Symmetric and Asymmetric Functions of Multiple Eigenvalues*

To illustrate the significance of symmetric functions of multiple eigenvalues in ensuring differentiability, the following example explores an illustrative case of such a function: $g(\lambda_1, \lambda_2, \lambda_3)$, i.e.,

$$g(\lambda_1, \lambda_2, \lambda_3) = \lambda_1^2 \lambda_2 \lambda_3 + \lambda_2^2 \lambda_3 \lambda_1 + \lambda_3^2 \lambda_1 \lambda_2 \tag{11}$$

Note that the function $g(\lambda_1, \lambda_2, \lambda_3)$ is invariant under any permutation of its arguments. So it is a symmetric function of the first eigen-cluster $\Lambda_1$ ($\Lambda_1 = \{\lambda_1, \lambda_2, \lambda_3\}$). The following function $h(\lambda_1, \lambda_2, \lambda_3)$ is constructed by omitting the product of $\lambda_2$ in the last term of Eq. (11). As a result, $h(\lambda_1, \lambda_2, \lambda_3)$ in Eq. (12) is not a symmetric function of $\Lambda_1$.

$$h(\lambda_1, \lambda_2, \lambda_3) = \lambda_1^2 \lambda_2 \lambda_3 + \lambda_2^2 \lambda_3 \lambda_1 + \lambda_3^2 \lambda_1 \tag{12}$$

To demonstrate the importance of symmetric functions of eigen-cluster in guaranteeing differentiability, the sensitivity analysis is performed in the case of $C_{5v}$ to accidental $I_h$ design variable symmetry. Based on the sensitivity results shown in Figure 9, it is concluded that only a symmetric function of the complete eigen-cluster can ensure differentiability under the condition of accidental symmetry (Figure 9(b)), whereas the asymmetric function of multiple eigenvalues is still not differentiable (Figure 9(a)).

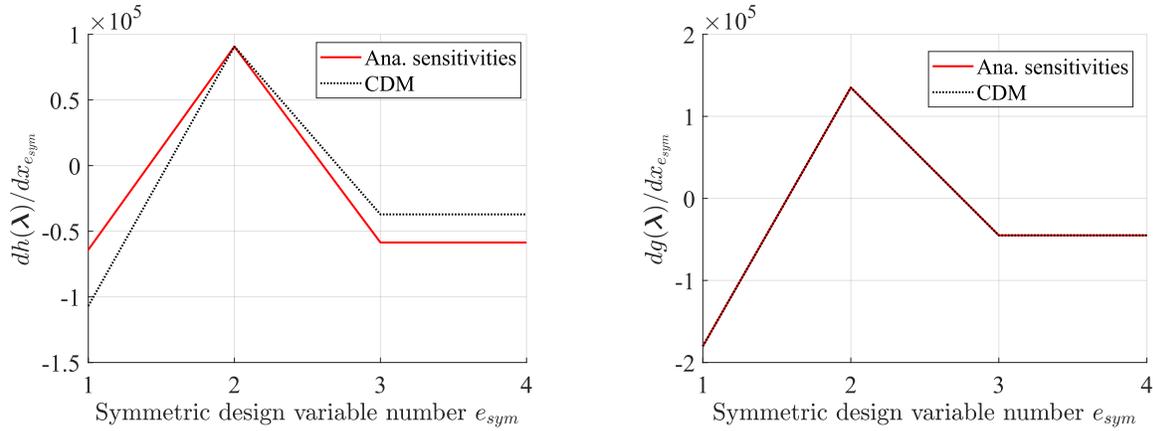

Figure 9. Asymmetric function (left) and symmetric function (right) sensitivities w.r.t all design variables ($C_{5v}$ to accidental $I_h$ design variable symmetry).

### 3.5 Icosahedral Trusses

The occurrence of accidental symmetry is further investigated in the example of icosahedral trusses, which have the same $I_h$ geometric symmetry as the dodecahedral trusses. The coordinates and dimensions of the icosahedral truss are detailed in Appendix A.5. The first 3 zero rigid-body mode-related eigenvalues are omitted. The illustrations of the trusses with 2 types of design variable symmetries are shown in Figure 10. The case of no-design variable symmetry to the accidental $I_h$ symmetry is examined. Although there is no enforced symmetry, all the design variables match those in the full $I_h$ symmetry to represent the case of accidental symmetry (Figure 29). Hence, the icosahedral truss with the accidental $I_h$ symmetry has the same eigenvalues as the truss with full $I_h$ design variable symmetry (Table 18).

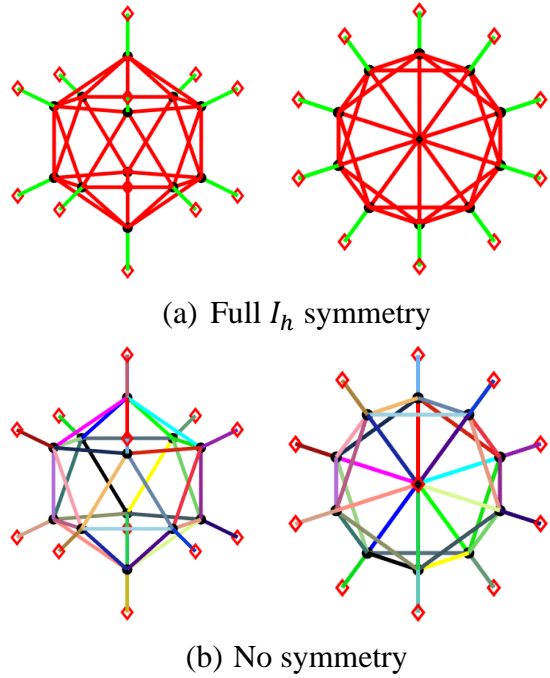

(a) Full $I_h$ symmetry

(b) No symmetry

Figure 10. Icosahedral trusses with 2 types of design variable symmetries (left: isometric view, right: plane view).

Table 18. Eigenanalysis results of icosahedral trusses with full or accidental $I_h$ design variable symmetry.

| | Full or acc. $I_h$ symmetry | | | | | | | |
|---|---|---|---|---|---|---|---|---|
| $\lambda_1$ | 30.772 | $\lambda_{10}$ | 150.000 | $\lambda_{19}$ | 230.181 | $\lambda_{28}$ | 612.055 |
| $\lambda_2$ | 30.772 | $\lambda_{11}$ | 150.000 | $\lambda_{20}$ | 230.181 | $\lambda_{29}$ | 612.055 |
| $\lambda_3$ | 30.772 | $\lambda_{12}$ | 150.000 | $\lambda_{21}$ | 514.058 | $\lambda_{30}$ | 625.354 |
| $\lambda_4$ | 67.765 | $\lambda_{13}$ | 153.165 | $\lambda_{22}$ | 514.058 | $\lambda_{31}$ | 625.354 |
| $\lambda_5$ | 67.765 | $\lambda_{14}$ | 153.165 | $\lambda_{23}$ | 514.058 | $\lambda_{32}$ | 625.354 |
| $\lambda_6$ | 67.765 | $\lambda_{15}$ | 153.165 | $\lambda_{24}$ | 514.058 | $\lambda_{33}$ | 630.311 |
| $\lambda_7$ | 67.765 | $\lambda_{16}$ | 153.165 | $\lambda_{25}$ | 612.055 | | |
| $\lambda_8$ | 67.765 | $\lambda_{17}$ | 153.165 | $\lambda_{26}$ | 612.055 | | |
| $\lambda_9$ | 150.000 | $\lambda_{18}$ | 230.181 | $\lambda_{27}$ | 612.055 | | |

In the case of $I_h$ design variable symmetry, the icosahedral truss has 3 clusters of invariant multiple eigenvalues with the values of 67.765, 150.00, and 514.058. Again, the eigen-clusters are constructed according to the eigen-analysis results recorded in Table 18: $\underbrace{\lambda_1 = \lambda_2 = \lambda_3}_{\Lambda_1} <$ $\underbrace{\lambda_4 = \lambda_5 = \lambda_6 = \lambda_7 = \lambda_8}_{\Lambda_2} < \underbrace{\lambda_9 = \lambda_{10} = \lambda_{11} = \lambda_{12}}_{\Lambda_3} < \underbrace{\lambda_{13} = \lambda_{14} = \lambda_{15} = \lambda_{16} = \lambda_{17}}_{\Lambda_4} <$ $\underbrace{\lambda_{18} = \lambda_{19} = \lambda_{20}}_{\Lambda_5} < \underbrace{\lambda_{21} = \lambda_{22} = \lambda_{23} = \lambda_{24}}_{\Lambda_6} < \underbrace{\lambda_{25} = \lambda_{26} = \lambda_{27} = \lambda_{28} = \lambda_{29}}_{\Lambda_7} <$ $\underbrace{\lambda_{30} = \lambda_{31} = \lambda_{32}}_{\Lambda_8} < \underbrace{\lambda_{33}}_{\Lambda_9}$. The eigenvalue sensitivity check results are summarized in Figure 11, where only the sensitivity of the first eigenvalue in each eigen-cluster is shown. The sensitivity results demonstrate that the individual multiple eigenvalues, without clustering, are not differentiable w.r.t the design variables in the case of accidental symmetry. Again, utilizing the concept of symmetric functions of multiple eigenvalues, the cluster means are differentiable w.r.t design variables (Figure 12).

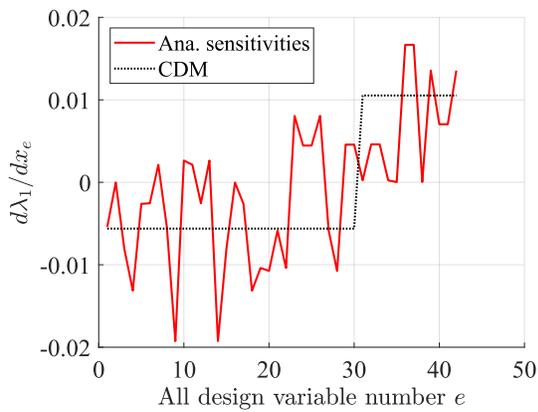
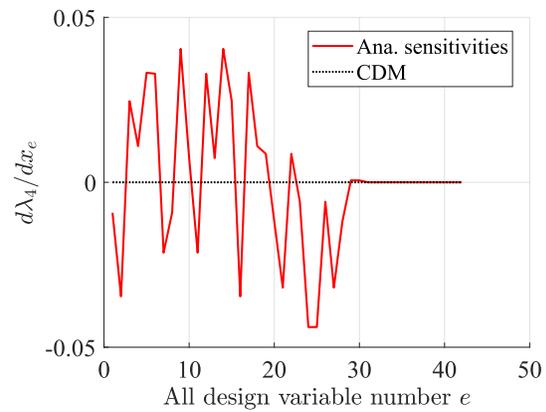

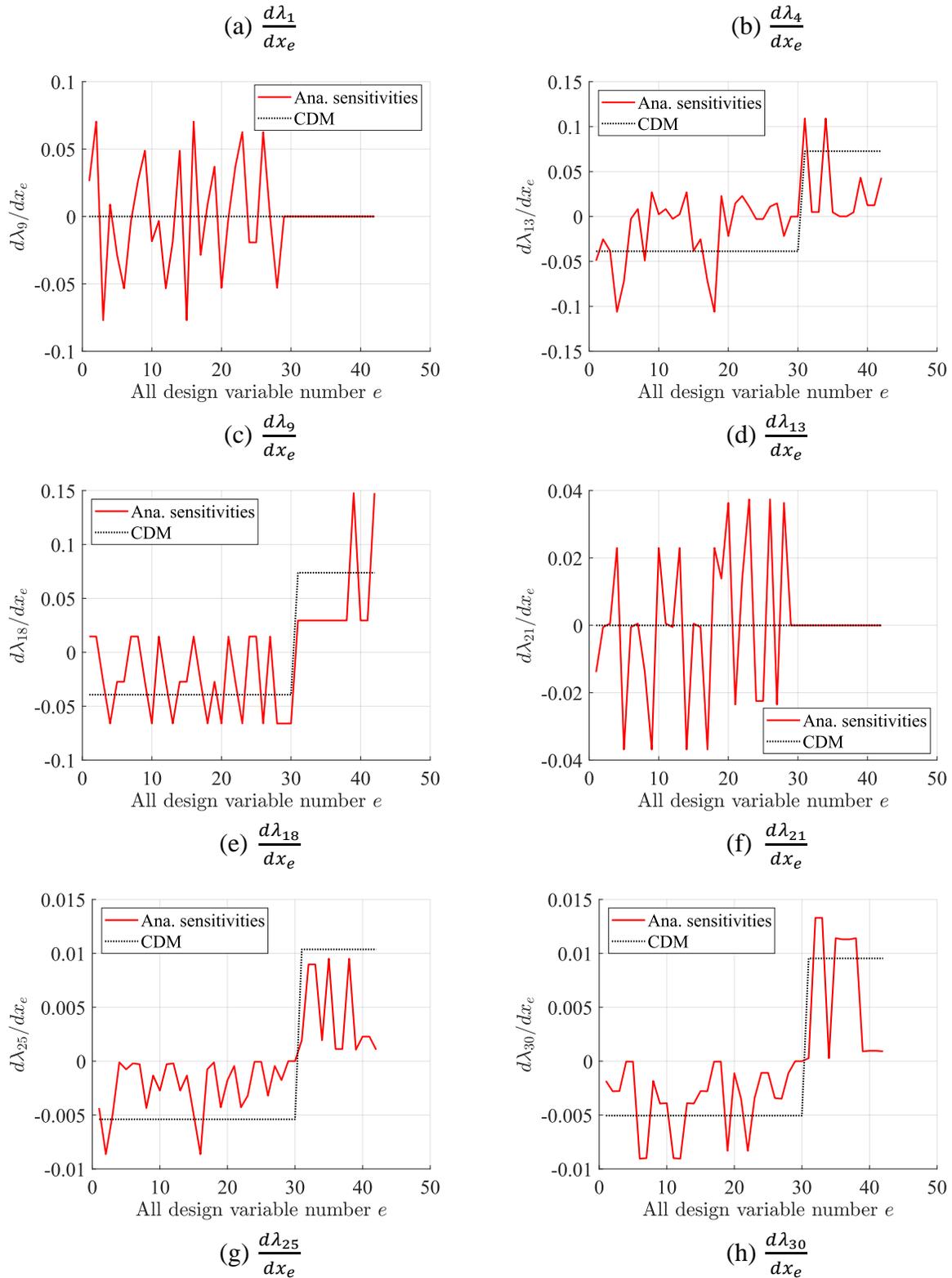

(a) $\frac{d\lambda_1}{dx_e}$

(b) $\frac{d\lambda_4}{dx_e}$

(c) $\frac{d\lambda_9}{dx_e}$

(d) $\frac{d\lambda_{13}}{dx_e}$

(e) $\frac{d\lambda_{18}}{dx_e}$

(f) $\frac{d\lambda_{21}}{dx_e}$

(g) $\frac{d\lambda_{25}}{dx_e}$

(h) $\frac{d\lambda_{30}}{dx_e}$

Figure 11. Multiple eigenvalue sensitivity check results of the icosahedral truss with no symmetry to accidental $I_h$ design variable symmetry.

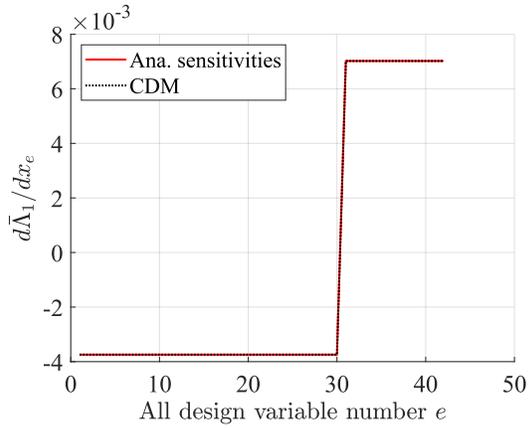

(a) $\dfrac{d\bar{\Lambda}_1}{dx_e}$

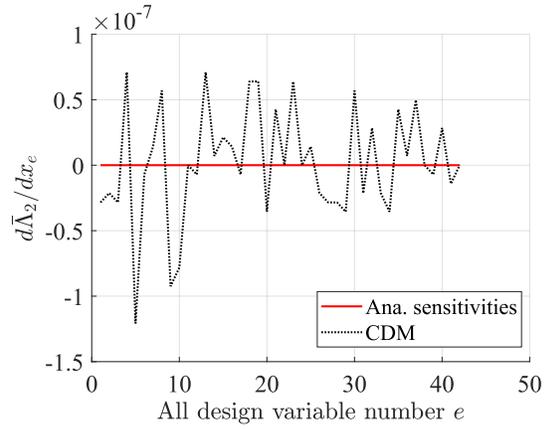

(b) $\dfrac{d\bar{\Lambda}_2}{dx_e}$

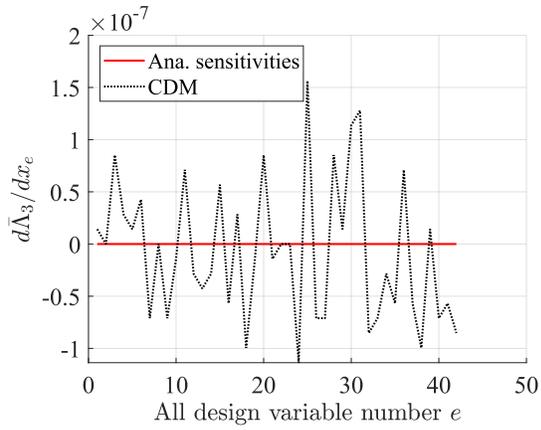

(c) $\dfrac{d\bar{\Lambda}_3}{dx_e}$

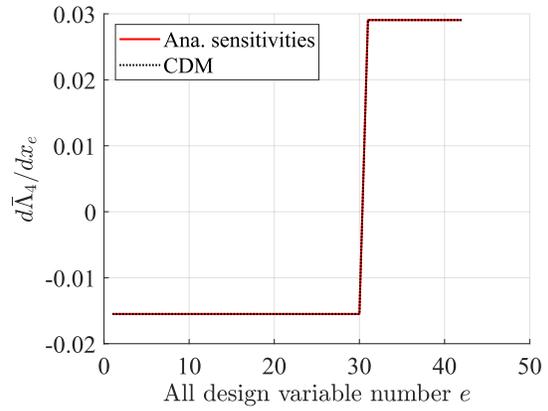

(d) $\dfrac{d\bar{\Lambda}_4}{dx_e}$

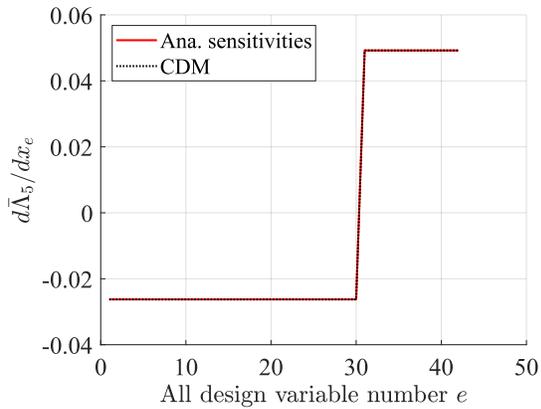

(e) $\dfrac{d\bar{\Lambda}_5}{dx_e}$

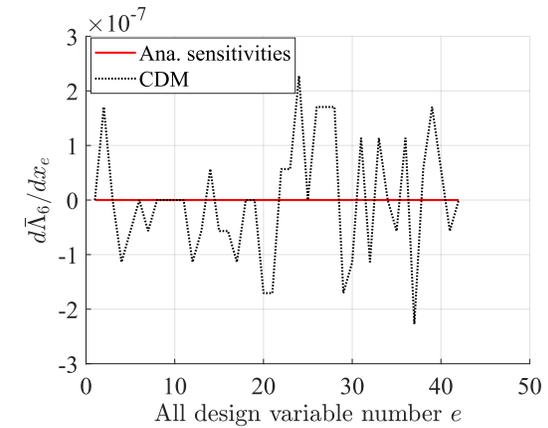

(f) $\dfrac{d\bar{\Lambda}_6}{dx_e}$

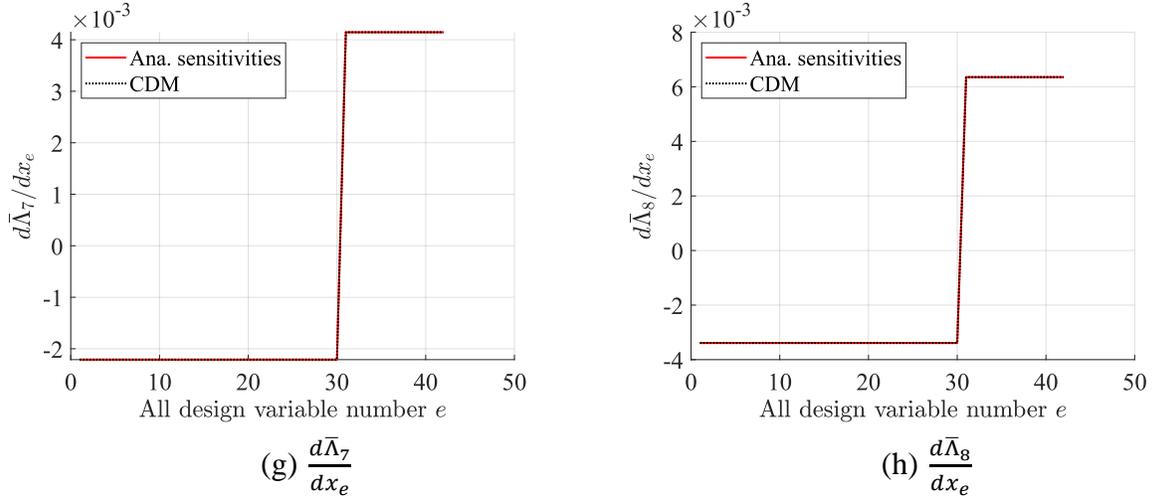

Figure 12. Mean functions of eigen-clusters with multiple eigenvalue sensitivity check results of the icosahedral truss with no symmetry to accidental $I_h$ design variable symmetry.

Aside from the cluster mean, the $p$-norm [42-45] and Kreisselmeier-Steinhauser (KS) functions [42, 46, 47] are also frequently used in eigenvalue optimization. The $p$-norm and KS functions are smooth and symmetric functions of eigenvalues. However, care should be taken in the construction of such smooth symmetric functions. These symmetric functions are differentiable only if complete eigen-clusters are utilized in their construction. This crucial point of *completeness* of the eigen-clusters has been overlooked in many past studies. This is also true for the mean function discussed above, where complete clusters are used. This section further clarifies the crucial role of complete eigen-clusters in ensuring the differentiability of $p$-norm and KS functions.

### 3.5.1 Symmetric Function: p-norm Function

The $p$-norm function, $\|\blacksquare\|_p$, can be utilized as a smooth approximation of the maximum norm $\|\blacksquare\|_\infty$. Using the $p$-norm function, the smooth approximation of the maximum value denoted as $\lambda^*$ of the first $n$ eigenvalues, $\boldsymbol{\lambda}$, is formulated as

$$\lambda^* = \|\boldsymbol{\lambda}\|_p = \left(\sum_{k=1}^{n} \lambda_k^p\right)^{1/p} \tag{13}$$

where $\lambda_{max} = \lim_{p \to \infty} \|\boldsymbol{\lambda}\|_p = \|\boldsymbol{\lambda}\|_\infty$.

When the complete eigen-clusters are contained in its arguments, the $p$-norm function is also a symmetric function of multiple eigenvalues. It should be noted that all repeated eigenvalues in the considered clusters must be included in the $p$-norm to be differentiable, i.e., the eigen-clusters must be complete. The following sensitivity results aim to demonstrate the necessity of the complete eigen-clusters in constructing differentiable symmetric functions of multiple eigenvalues. Figure 13 compares the sensitivity results for two cases, i.e., with and without the inclusion of complete eigen-clusters in constructing $p$-norm. Figure 13 (left) illustrates the sensitivity check results of the $p$-norm of the first 15 eigenvalues in Table 18, where the 4th cluster $\Lambda_4$ is incomplete. The mismatch between the CDM and analytical sensitivities indicates that the $p$-norm function of the incomplete cluster is not differentiable. However, by including the first 17 eigenvalues, $\Lambda_4$ is now complete, and the corresponding $p$-norm function becomes differentiable (Figure 13 (right)).

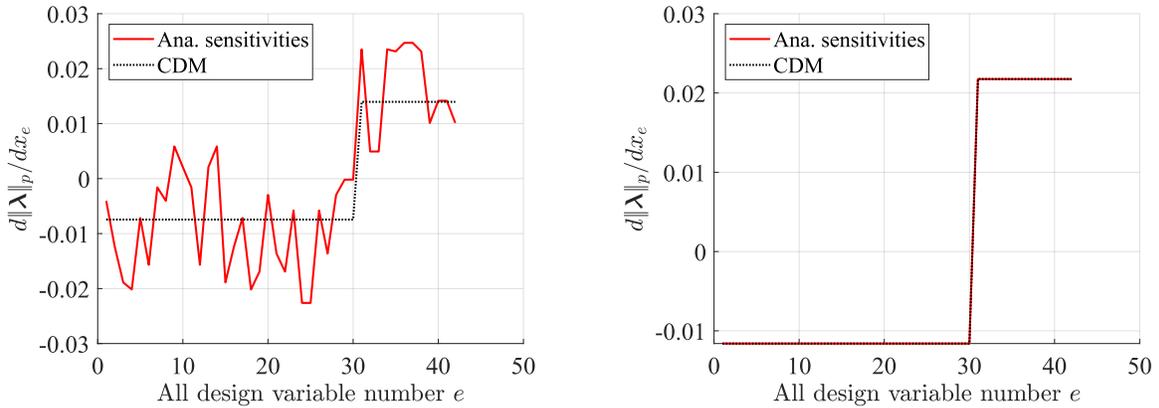

Figure 13. $p$-norm sensitivities w.r.t all design variables ($p = 10$) (no symmetry to accidental $I_h$ design variable symmetry) when the first 4 eigen-clusters are incomplete (left) and complete (right).

### 3.5.2 Symmetric Function: KS Function

Another commonly used smooth approximation of the maximum of the first $n$ eigenvalues is the KS function and is expressed as

$$\lambda^* = KS(\lambda, q) = \frac{1}{q}\ln\left(\sum_{k=1}^{n} \exp(q\lambda_k)\right) \tag{14}$$

and $\lambda_{max} = \lim_{q \to \infty} \lambda^*$.

To again illustrate the importance of completeness, Figure 14 shows the KS function sensitivity results when the first 15 eigenvalues (cluster $\Lambda_4$ is incomplete) and the first 17 eigenvalues (cluster $\Lambda_4$ is complete) are used in the KS function. This result again demonstrates that the KS function is differentiable only when all repeated eigenvalues in the considered eigen-clusters are included.

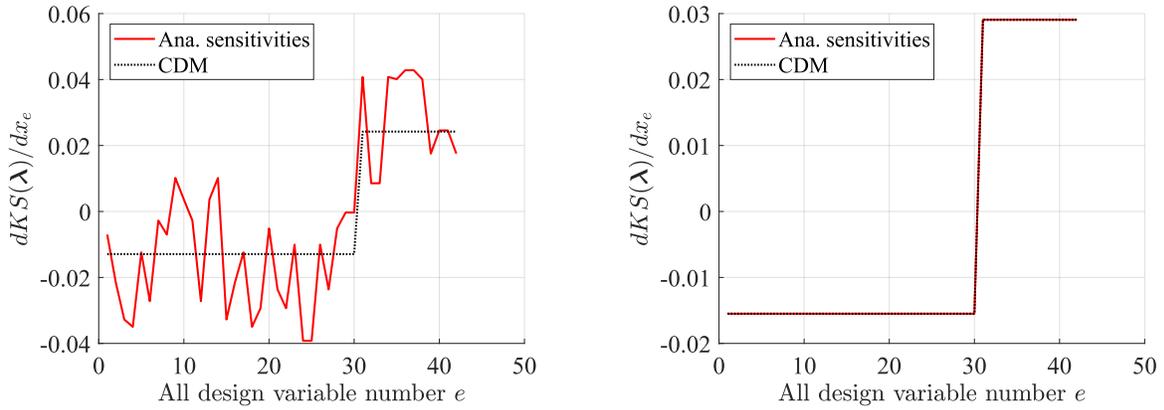

Figure 14. KS function sensitivities w.r.t all design variables ($q = 10$) (no symmetry to accidental $I_h$ design variable symmetry) when the first 4 eigen-clusters are incomplete (left) and complete (right).

### 3.5.3 Differentiable Function - General Case

A general method of constructing a *differentiable* function of $m$-clusters (with indexes $k_1, k_2 ..., k_m$) of multiple eigenvalues is to express it in terms of eigen-cluster means, i.e., a smooth function $f: \mathbb{R}^m \to \mathbb{R}$ expressed as $f(\overline{\boldsymbol{\Lambda}}) = f(\overline{\Lambda}_{k_1}, \overline{\Lambda}_{k_2}, ..., \overline{\Lambda}_{k_m})$, is *differentiable*. This is because the cluster-mean function is differentiable, and any smooth function composed with it remains symmetric with respect to individual eigen-clusters, and thus differentiable. Following is an example of such a function:

$$f(\overline{\Lambda}_1, \overline{\Lambda}_4, \overline{\Lambda}_5) = (\overline{\Lambda}_1)^2 * \overline{\Lambda}_4 + 100 * \sin(\overline{\Lambda}_4 + \overline{\Lambda}_5) \tag{15}$$

where $\overline{\Lambda}_1, \overline{\Lambda}_4$ and $\overline{\Lambda}_5$ are the cluster means of 3 non-invariant eigen-clusters with multiplicities greater than 1 in the case of no symmetry to accidental $I_h$ design variable symmetry (Table 18). Note that the function $f(\overline{\Lambda}_1, \overline{\Lambda}_4, \overline{\Lambda}_5)$ need not be symmetric with respect to its arguments to ensure differentiability. The sensitivity check result shown in Figure 15 demonstrates that such a smooth function is differentiable.

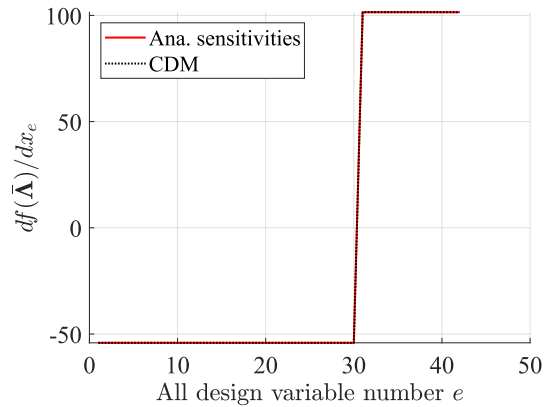

Figure 15. Smooth function of eigen-cluster means sensitivities w.r.t all design variables (no symmetry to accidental $I_h$ design variable symmetry).

## 4 Conclusions

In this work, the multiplicities of eigenvalues and their differentiability under various symmetry conditions are numerically investigated in the context of eigenfrequency problems. To this end, 3-D truss structures with different geometric and design variable symmetries are studied. The results show that the eigenvalues can be classified as simple and multiple (repeated). It is shown that *multiple* eigenvalues only arise when there is underlying symmetry in the structure. This finding is consistent with the results by [15], which utilized a group-theoretic approach to symmetry and eigenvalue multiplicity. Furthermore, the multiplicity of repeated eigenvalues decreases with the

Page 40 of 58

reduction in symmetry. Using the numerical examples, the differentiability of multiple eigenvalues is also studied, and the findings are summarized as follows:

(a) Multiple *invariant* eigenvalues are trivially differentiable.

(b) When reduced design variable symmetry is enforced, multiple eigenvalues are not differentiable w.r.t symmetric design variables in the case of *accidental* symmetry. The accidental symmetry increases the underlying symmetry of the problem, and the multiple eigenvalues are not Fréchet differentiable in this case.

(c) The differentiability of smooth symmetric functions of multiple eigenvalues is also investigated. It is shown that to ensure differentiability, the complete eigen-clusters should be included in the construction of the smooth symmetric functions such as the mean, $p$-norm and KS functions. These functions are then differentiable even in the case of accidental symmetry, where multiple eigenvalues are not differentiable. Furthermore, a method for constructing differentiable functions of eigenvalues using eigen-clusters is presented. These findings resolve the ambiguities surrounding the differentiability of multiple eigenvalues.

For practical considerations in eigenvalue optimization problems, it is emphasized that symmetric functions of complete eigen-clusters need to be constructed to ensure differentiability. Such treatment is required when accidental symmetry occurs at a design point because the multiple eigenvalues are not differentiable at such a design point. Although the results are presented in the context of natural frequency analysis, similar considerations apply to other structural eigenvalue problems, including stability [27, 41, 43, 48]. Furthermore, this work only considers numerical studies on 3-D trusses; work is underway to investigate this issue via group theoretical tools.

**Acknowledgments**

The presented work is supported in part by the U.S. Department of Energy by Lawrence Livermore Laboratory under Contract #B652057 and by the University of Notre Dame. Any opinions, findings, conclusions, and recommendations expressed in this article are those of the authors and do not necessarily reflect the views of the sponsors.

**Conflict of Interest**

On behalf of all authors, the corresponding author states that there is no conflict of interest.

**Replication of Results**

The source code used to generate these results is shared by the authors on GitHub [38].

# Appendix A

To assist readers in replicating the study's results, this appendix provides detailed information on the coordinates of all the truss structures presented in this work. Additionally, the assignment and values of the design variables across all examples are summarized in this appendix. Grayscale versions of the figures are provided after their colored counterparts.

## A.1 Truss Dome

The truss domes are formed by equally dividing the center radially into $N$ subsections. For illustration purposes, the dimension information of the dome truss with $N = 8$ is shown in Figure 16. Based on this example, other truss domes with $N = 3, 4, 5, 6$ and 7 can be constructed.

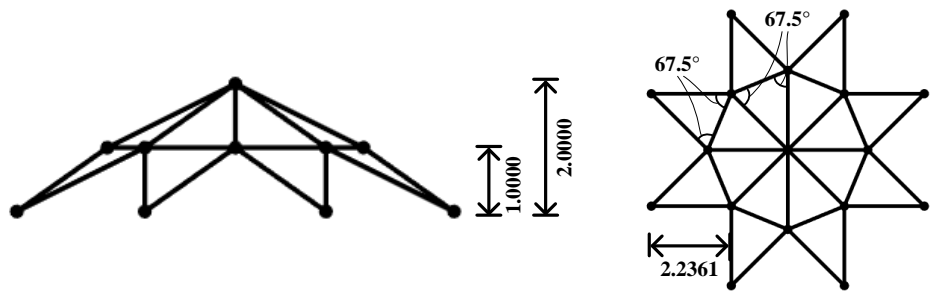

Figure 16. Dome truss ($N = 8$) dimensions (left: elevation view, right: plane view).

The data of the design variables of the truss dome ($N = 8$) with 2 types of symmetry conditions is summarized in Figure 17. In the no-design variable symmetry condition, the design variables $x_1$ to $x_{32}$ are assigned values of 100 to 410 with increments of 10 (Figure 17(a)). The idea behind the design variable assignment is the same for truss dome with $N = 3,4,5,6,7$ and with no-design variable symmetry. In the $C_{8v}$ symmetry case, the symmetric design variables are grouped by colors, and their values are summarized in Figure 17(b). The categorization and the values of symmetric variables in the $C_{Nv}$ symmetry cases remain the same across all values of $N$.

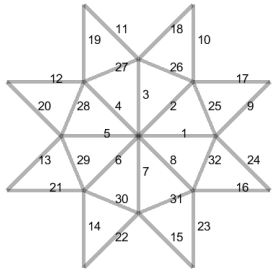 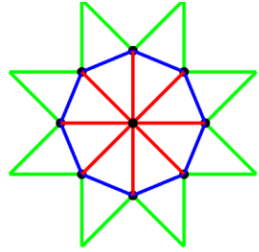

| No design var. symmetry | |
|---|---|
| Design var. | Value |
| $x_1$ | 100 |
| $x_2$ | 110 |
| $\vdots$ | $\vdots$ |
| $x_{32}$ | 410 |

| $C_{8v}$ design var. symmetry | |
|---|---|
| Design var. | Value |
| $x_{1_{sym}}$ (red) | 150 |
| $x_{2_{sym}}$ (green) | 200 |
| $x_{3_{sym}}$ (blue) | 50 |

(a) No-design variable symmetry  (b) $C_{8v}$ design variable symmetry

Figure 17. Truss dome design variable summary.

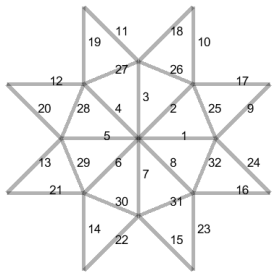 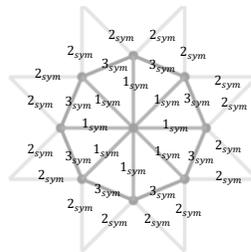

| No design var. symmetry | |
|---|---|
| Design var. | Value |
| $x_1$ | 100 |
| $x_2$ | 110 |
| $\vdots$ | $\vdots$ |
| $x_{32}$ | 410 |

| $C_{8v}$ design var. symmetry | |
|---|---|
| Design var. | Value |
| $x_{1_{sym}}$ | 150 |
| $x_{2_{sym}}$ | 200 |
| $x_{3_{sym}}$ | 50 |

(a) No-design variable symmetry  (b) $C_{8v}$ design variable symmetry

Figure 18. Truss dome design variable summary (grayscale).

### *A.2 Tetrahedral Truss*

In the center of the tetrahedral truss is a tetrahedron whose all four faces are equilateral triangles (Figure 6). Extending outward from each of the 4 vertices of the tetrahedron are 4 truss elements with pin supports at the other ends (Figure 6). The dimensions of the tetrahedral truss are detailed in Figure 19. The values of the design variables of the tetrahedral truss with 3 types of symmetry cases are summarized in Figure 20.

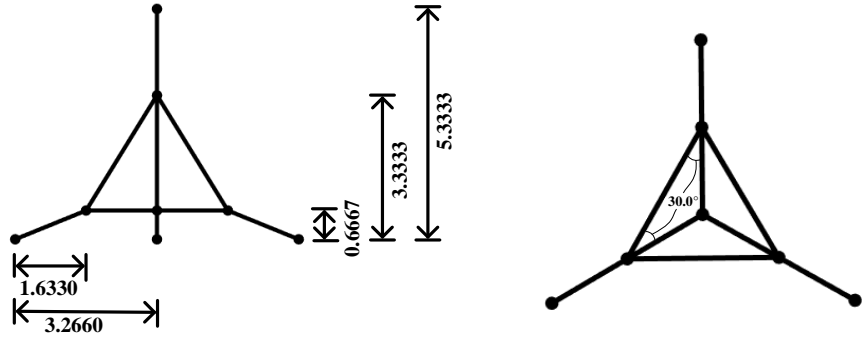

Figure 19. Tetrahedral truss dimensions (left: elevation view, right: plane view).

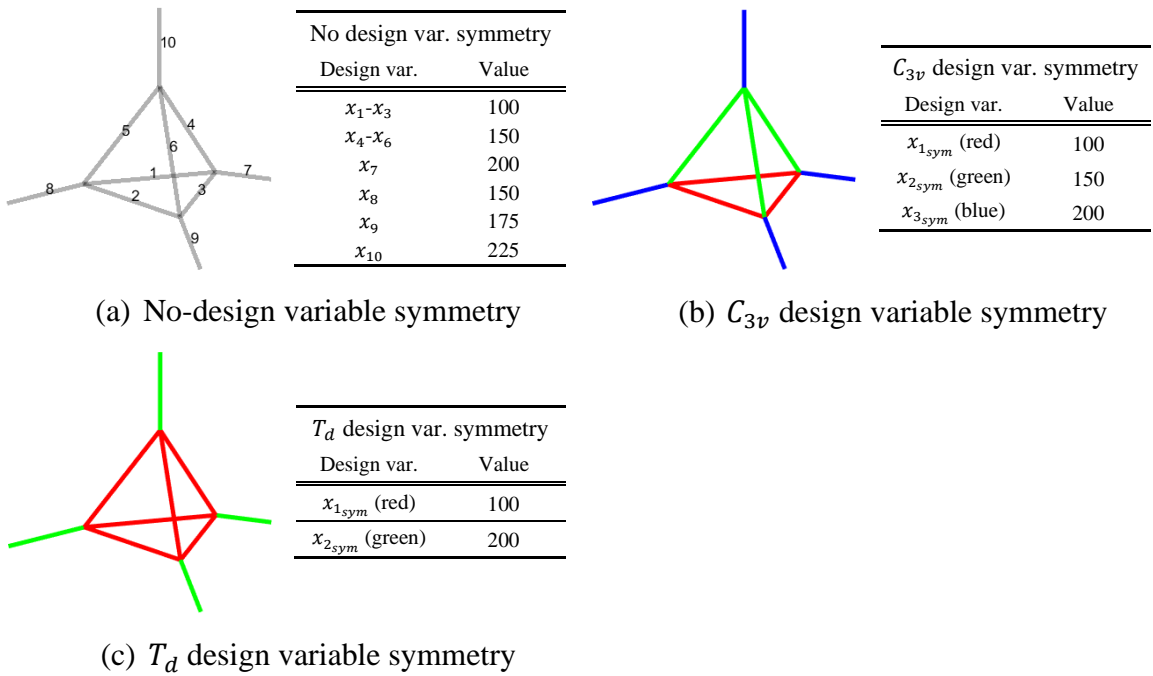

(a) No-design variable symmetry

(b) $C_{3v}$ design variable symmetry

(c) $T_d$ design variable symmetry

Figure 20. Tetrahedral truss design variable summary.

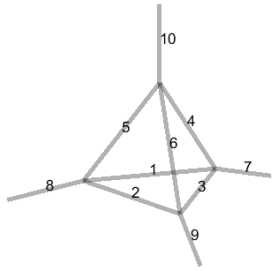

| No design var. symmetry | |
|---|---|
| Design var. | Value |
| $x_1$-$x_3$ | 100 |
| $x_4$-$x_6$ | 150 |
| $x_7$ | 200 |
| $x_8$ | 150 |
| $x_9$ | 175 |
| $x_{10}$ | 225 |

(a) No-design variable symmetry

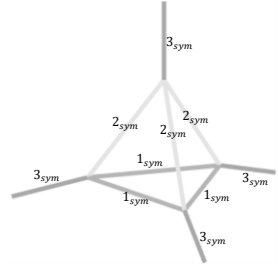

| $C_{3v}$ design var. symmetry | |
|---|---|
| Design var. | Value |
| $x_{1_{sym}}$ | 100 |
| $x_{2_{sym}}$ | 150 |
| $x_{3_{sym}}$ | 200 |

(b) $C_{3v}$ design variable symmetry

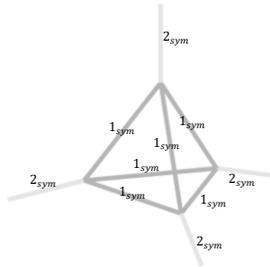

| $T_d$ design var. symmetry | |
|---|---|
| Design var. | Value |
| $x_{1_{sym}}$ | 100 |
| $x_{2_{sym}}$ | 200 |

(c) $T_d$ design variable symmetry

Figure 21. Tetrahedral truss design variable summary (grayscale).

## A.3 Octahedral Truss

The octahedral truss is structured around a central regular octahedron, which consists of 12 members. Additionally, it includes 6 outer elements extending from the vertices of the octahedron, each supported by pins at their ends (Figure 7). The dimensions of the octahedral truss are shown in Figure 22. The information regarding the assignment and values of the design variables is summarized in Figure 23.

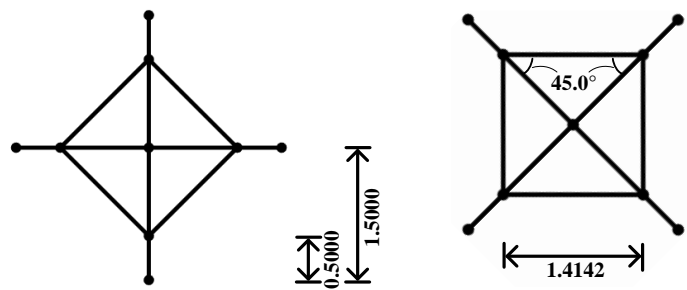

Figure 22. Octahedral truss dimensions (left: elevation view, right: plane view).

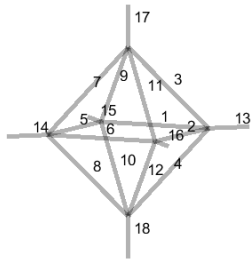

| No design var. symmetry | |
| --- | --- |
| Design var. | Value |
| $x_1$ | 100 |
| $x_2$ | 110 |
| ⋮ | ⋮ |
| $x_{18}$ | 270 |

(a) No design variable symmetry

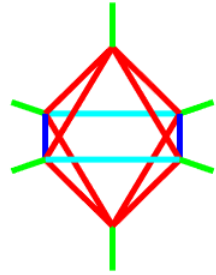

| $C_{2v}$ design var. symmetry | |
| --- | --- |
| Design var. | Value |
| $x_{1_{sym}}$ (red) | 150 |
| $x_{2_{sym}}$ (green) | 200 |
| $x_{3_{sym}}$ (blue) | 150 |
| $x_{4_{sym}}$ (cyan) | 175 |

(b) $C_{2v}$ design variable symmetry

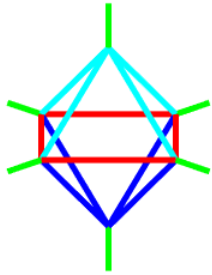

| $C_{4v}$ design var. symmetry | |
| --- | --- |
| Design var. | Value |
| $x_{1_{sym}}$ (red) | 150 |
| $x_{2_{sym}}$ (green) | 300 |
| $x_{3_{sym}}$ (blue) | 225 |
| $x_{4_{sym}}$ (cyan) | 250 |

(c) $C_{4v}$ design variable symmetry

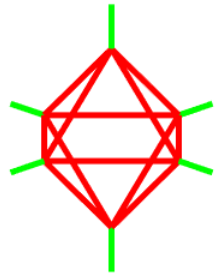

| $O_h$ design var. symmetry | |
| --- | --- |
| Design var. | Value |
| $x_{1_{sym}}$ (red) | 150 |
| $x_{2_{sym}}$ (green) | 200 |

(d) $O_h$ design variable symmetry

Figure 23. Octahedral truss design variable symmetry.

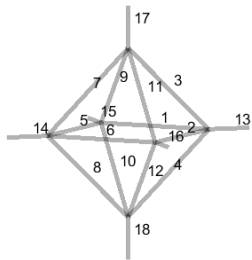

| No design var. symmetry | |
| --- | --- |
| Design var. | Value |
| $x_1$ | 100 |
| $x_2$ | 110 |
| ⋮ | ⋮ |
| $x_{18}$ | 270 |

(a) No design variable symmetry

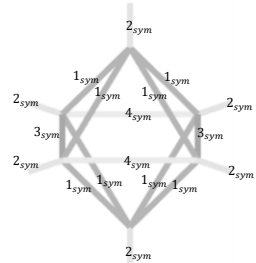

| $C_{2v}$ design var. symmetry | |
| --- | --- |
| Design var. | Value |
| $x_{1_{sym}}$ (red) | 150 |
| $x_{2_{sym}}$ (green) | 200 |
| $x_{3_{sym}}$ (blue) | 150 |
| $x_{4_{sym}}$ (cyan) | 175 |

(b) $C_{2v}$ design variable symmetry

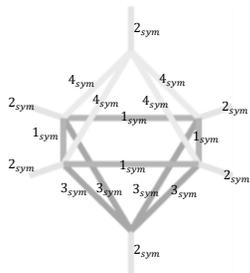

| $C_{4v}$ design var. symmetry | |
| --- | --- |
| Design var. | Value |
| $x_{1_{sym}}$ | 150 |
| $x_{2_{sym}}$ | 300 |
| $x_{3_{sym}}$ | 225 |
| $x_{4_{sym}}$ | 250 |

(c) $C_{4v}$ design variable symmetry

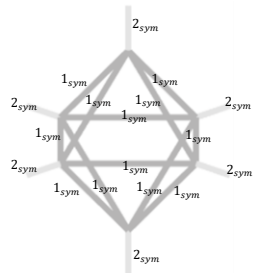

| $O_h$ design var. symmetry | |
| --- | --- |
| Design var. | Value |
| $x_{1_{sym}}$ | 150 |
| $x_{2_{sym}}$ | 200 |

(d) $O_h$ design variable symmetry

Figure 24. Octahedral truss design variable symmetry (grayscale).

## *A.4 Dodecahedral Truss*

The dodecahedral truss centers around a regular dodecahedron which is composed of 12 regular pentagonal faces, 3 meeting at each vertex. The 20 vertices of a regular dodecahedron, centered at the origin, are defined by the Cartesian coordinates shown in Figure 25(a). The dodecahedral truss also consists of truss elements extending from each vertex of the center dodecahedron, which are supported by pins at the other ends (Figure 8). The coordinates of the support nodes are summarized in Figure 25(b).

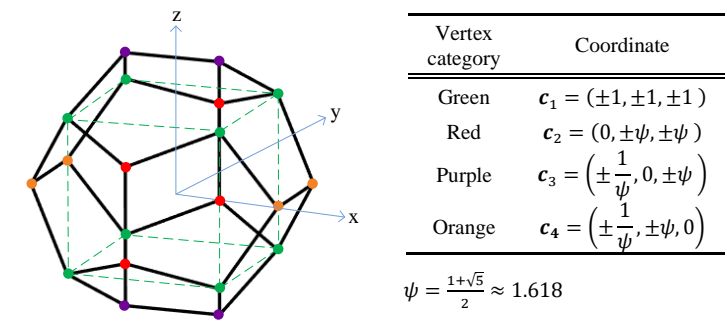

| Vertex category | Coordinate |
| --- | --- |
| Green | $c_1 = (\pm 1, \pm 1, \pm 1)$ |
| Red | $c_2 = (0, \pm\psi, \pm\psi)$ |
| Purple | $c_3 = \left(\pm\frac{1}{\psi}, 0, \pm\psi\right)$ |
| Orange | $c_4 = \left(\pm\frac{1}{\psi}, \pm\psi, 0\right)$ |

$\psi = \frac{1+\sqrt{5}}{2} \approx 1.618$

(a) Dodecahedral truss vertex coordinates

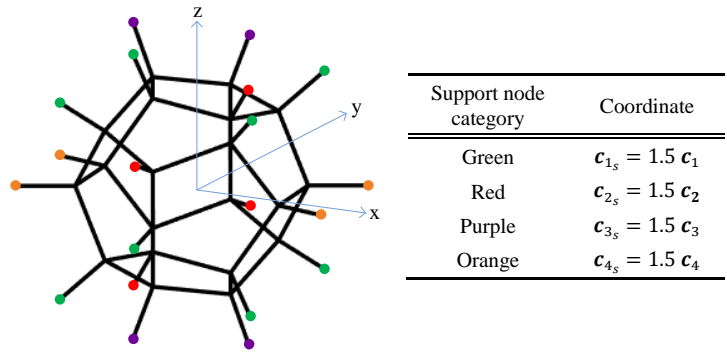

| Support node category | Coordinate |
| --- | --- |
| Green | $c_{1_s} = 1.5\, c_1$ |
| Red | $c_{2_s} = 1.5\, c_2$ |
| Purple | $c_{3_s} = 1.5\, c_3$ |
| Orange | $c_{4_s} = 1.5\, c_4$ |

(b) Dodecahedral truss support nodal coordinates

Figure 25. Dodecahedral truss nodal coordinates.

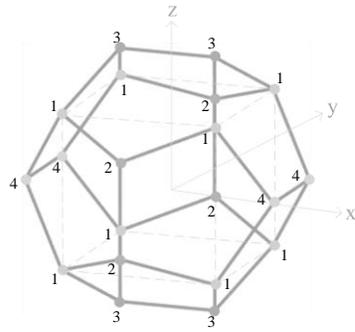

| Vertex category | Coordinate |
|---|---|
| 1 | $c_1 = (\pm 1, \pm 1, \pm 1)$ |
| 2 | $c_2 = (0, \pm\psi, \pm\psi)$ |
| 3 | $c_3 = \left(\pm\frac{1}{\psi}, 0, \pm\psi\right)$ |
| 4 | $c_4 = \left(\pm\frac{1}{\psi}, \pm\psi, 0\right)$ |

$$\psi = \frac{1+\sqrt{5}}{2} \approx 1.618$$

(a) Dodecahedral truss vertex coordinates

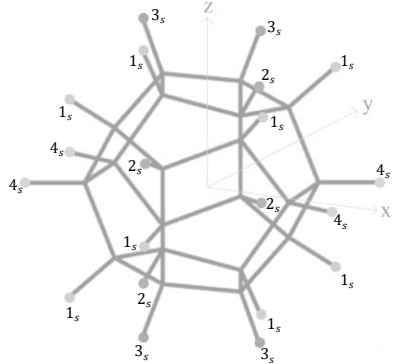

| Support node category | Coordinate |
|---|---|
| $1_s$ | $c_{1_s} = 1.5\, c_1$ |
| $2_s$ | $c_{2_s} = 1.5\, c_2$ |
| $3_s$ | $c_{3_s} = 1.5\, c_3$ |
| $4_s$ | $c_{4_s} = 1.5\, c_4$ |

(b) Dodecahedral truss support nodal coordinates

Figure 26. Dodecahedral truss nodal coordinates (grayscale).

### *A.5 Icosahedral Truss*

The icosahedral truss is structured around a central regular icosahedron, featuring 20 equilateral triangular faces with 5 faces meeting at each vertex. These vertices, defining the icosahedron centered at the origin, are specified by Cartesian coordinates illustrated in Figure 27(a). Truss elements extending outward from each vertex of this central icosahedron, are supported by pins at their opposite ends (Figure 10). The coordinates for these support nodes are detailed in Figure 27(b).

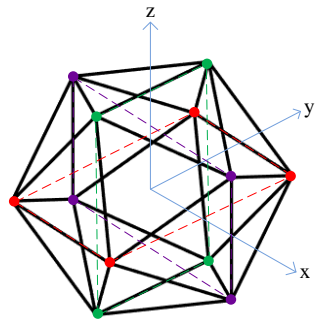

| Vertex category | Coordinate |
| --- | --- |
| Green | $c_1 = (0, \pm 1, \pm \psi)$ |
| Red | $c_2 = (\pm 1, \pm \psi, 0)$ |
| Purple | $c_3 = (\pm \psi, 0, \pm 1)$ |

$\psi = \frac{1+\sqrt{5}}{2} \approx 1.618$

(a) Icosahedral truss vertex coordinates

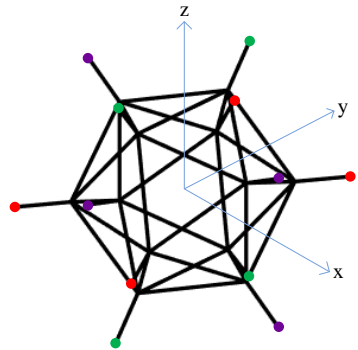

| Support node category | Coordinate |
| --- | --- |
| Green | $c_{1_s} = 1.5\, c_1$ |
| Red | $c_{2_s} = 1.5\, c_2$ |
| Purple | $c_{3_s} = 1.5\, c_3$ |

(b) Icosahedral truss support nodal coordinates

Figure 27. Icosahedral truss nodal coordinates.

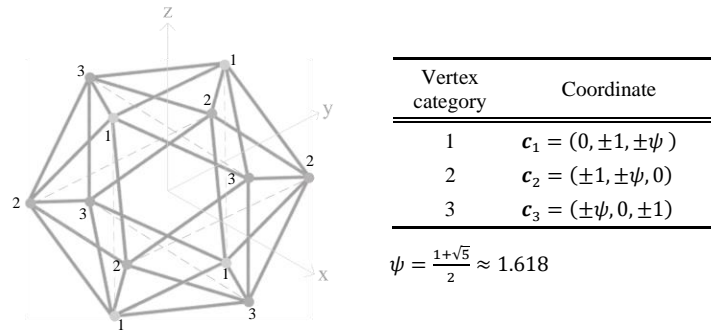

(a) Icosahedral truss vertex coordinates

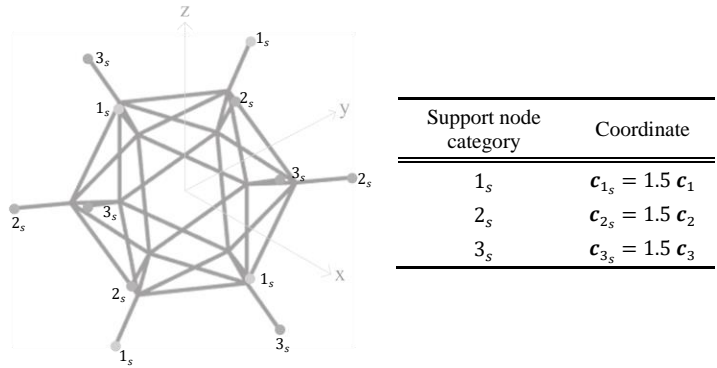

(b) Icosahedral truss support nodal coordinates

Figure 28. Icosahedral truss nodal coordinates (grayscale).

The design variable values of the two symmetry conditions of the icosahedral truss are summarized in Figure 29. The values of the design variables are specifically allocated to ensure the condition of no symmetry to accidental $I_h$ symmetry Figure 29(a). Therefore, the values of the element cross-sectional areas of the icosahedral truss are the same between the cases illustrated in Figure 29(a) and Figure 29(b). The distinction lies in the absence of enforced symmetry in the no symmetry to accidental $I_h$ symmetry case, whereas full symmetry is enforced in the $I_h$ design variable symmetry case.

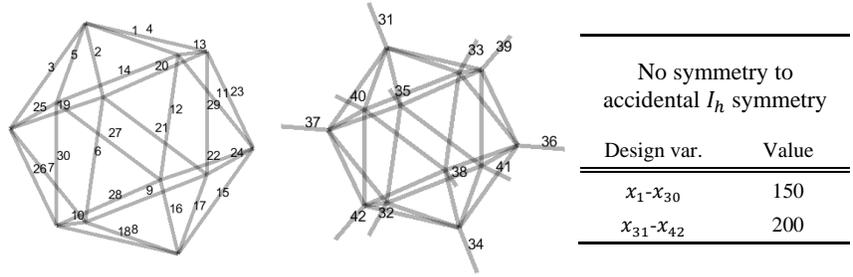

(a) No symmetry to accidental $I_h$ design variable symmetry

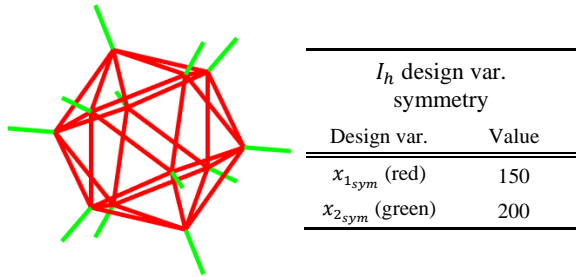

(b) $I_h$ design variable symmetry

Figure 29. Icosahedral truss design variable summary.

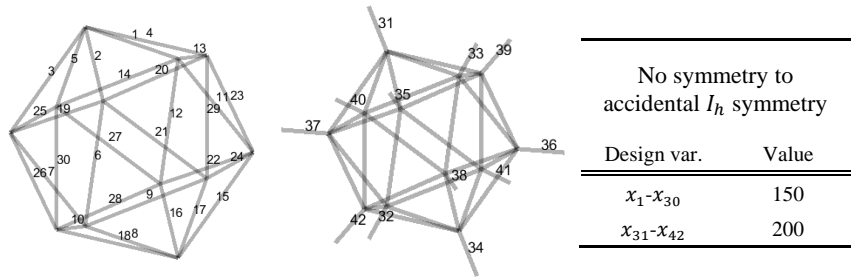

(a) No symmetry to accidental $I_h$ design variable symmetry

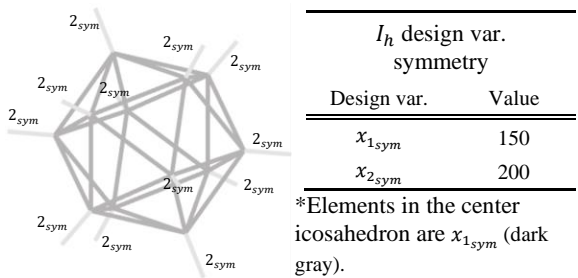

(b) $I_h$ design variable symmetry

Figure 30. Icosahedral truss design variable summary (grayscale).

20. Ikeda, K., I. Ario, and K. Torii, *Block-diagonalization analysis of symmetric plates.* International journal of solids and structures, 1992. **29**(22): p. 2779-2793.

21. Wohlever, J. and T. Healey, *A group theoretic approach to the global bifurcation analysis of an axially compressed cylindrical shell.* Computer Methods in Applied Mechanics and Engineering, 1995. **122**(3-4): p. 315-349.

22. Hughes, T.J., *The finite element method: linear static and dynamic finite element analysis*. 2012: Courier Corporation.

23. Bathe, K.-J., *Finite element procedures*. 2006: Klaus-Jurgen Bathe.

24. Kiyohiro, I., M. Kazuo, and F. Hiroshi, *Bifurcation hierarchy of symmetric structures.* International Journal of Solids and Structures, 1991. **27**(12): p. 1551-1573.

25. Kosaka, I. and C.C. Swan, *A symmetry reduction method for continuum structural topology optimization.* Computers & Structures, 1999. **70**(1): p. 47-61.

26. Du, J. and N. Olhoff, *Topological design of freely vibrating continuum structures for maximum values of simple and multiple eigenfrequencies and frequency gaps.* Structural and Multidisciplinary Optimization, 2007. **34**: p. 91-110.

27. Zhang, G., K. Khandelwal, and T. Guo, *Finite strain topology optimization with nonlinear stability constraints.* Computer Methods in Applied Mechanics and Engineering, 2023. **413**: p. 116119.

28. Banh, T.T., et al., *A robust dynamic unified multi-material topology optimization method for functionally graded structures.* Structural and Multidisciplinary Optimization, 2023. **66**(4): p. 75.